\documentclass[a4paper,fleqn]{cas-dc}

\usepackage[numbers]{natbib}
\usepackage{setspace}
\usepackage{wrapfig}
\usepackage{graphicx}
\usepackage{enumitem}
\usepackage{multirow}
\usepackage[final]{pdfpages}

\setlist[enumerate]{leftmargin=*}
\setlist[itemize]{leftmargin=*}

\newcommand{\V}[1]{\boldsymbol{#1}}
\newcommand{\F}[1]{\mathbf{#1}}

\begin{document}
\let\WriteBookmarks\relax
\def\floatpagepagefraction{1}
\def\textpagefraction{.001}
\setlength{\abovedisplayskip}{5pt}
\setlength{\belowdisplayskip}{5pt}

\includepdf[pages={1}]{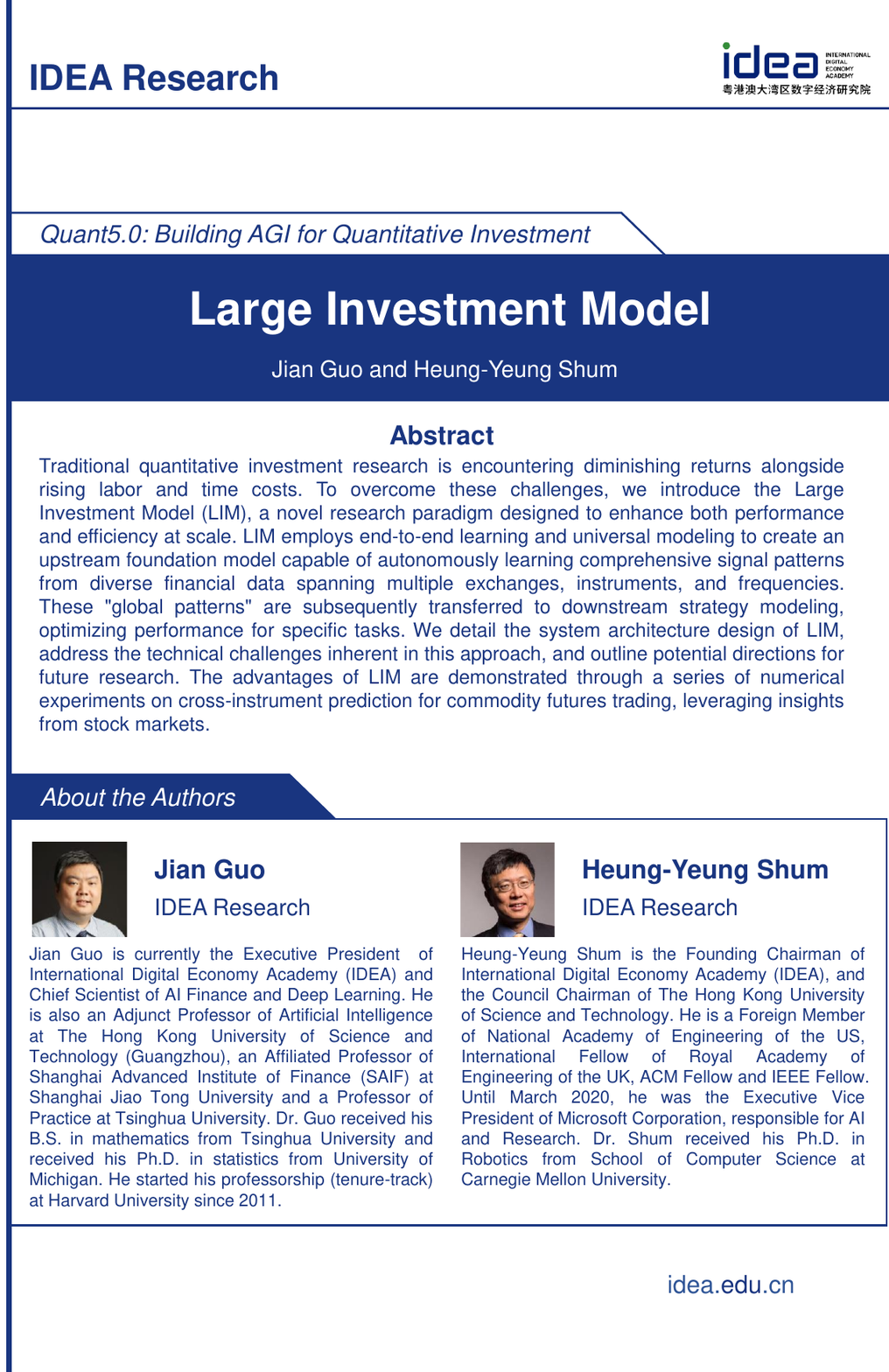}
\newpage

\shorttitle{Large Investment Model}

\newpage
\renewcommand*\listfigurename{\Large \centering List of Figures}
\renewcommand*\contentsname{\Large Table of Contents}
\tableofcontents
\newpage
\setcounter{page}{1}
\title [mode = title]{Large Investment Model}                      



%

\author{Jian Guo}
[orcid=0009-0003-5046-2588]
\cormark[1]
\ead{guojian@idea.edu.cn}



\author{Heung-Yeung Shum}
\ead{hshum@idea.edu.cn}

\affiliation{organization={IDEA Research, International Digital Economy Academy},
    city={Shenzhen},
    postcode={518045}, 
    country={China}}



\cortext[cor1]{Corresponding author}




\begin{abstract}
Traditional quantitative investment research is encountering diminishing returns alongside rising labor and time costs. To overcome these challenges, we introduce the Large Investment Model (LIM), a novel research paradigm designed to enhance both performance and efficiency at scale. LIM employs end-to-end learning and universal modeling to create an upstream foundation model capable of autonomously learning comprehensive signal patterns from diverse financial data spanning multiple exchanges, instruments, and frequencies. These ``global patterns'' are subsequently transferred to downstream strategy modeling, optimizing performance for specific tasks. We detail the system architecture design of LIM, address the technical challenges inherent in this approach, and outline potential directions for future research. The advantages of LIM are demonstrated through a series of numerical experiments on cross-instrument prediction for commodity futures trading, leveraging insights from stock markets.
\end{abstract}

\begin{keywords}

Artificial General Intelligence \sep End-to-End \sep Large Investment Model \sep Quantitative Investment \sep Foundation Model \sep Multimodal Large Language Model

\end{keywords}

\maketitle


\section{Introduction}
Quantitative investment (quant) involves financial investment strategies driven by mathematical, statistical, or machine learning models, and it uses powerful computers execute trading instructions derived from quant models at speeds and frequencies unattainable by human traders. In particular, deep learning techniques are widely applied in quant modeling, such as stock/futures trend prediction~\cite{zhang_stock_2017,xu_stock_2018,hu_listening_2018,feng_time_2021}, stock selection~\cite{feng_temporal_2019,sawhney_spatiotemporal_2020,sawhney_stock_2021}, portfolio optimization~\cite{jiang_deep_2017,wang_alphastock_2019,wang_deeptrader_2021,zhang_cost-sensitive_2022,liu_deep_2023} and algorithmic trading~\cite{lin_end2end_2020,fang_universal_2021,sun_deepscalper_2022,fang_learning_2023,qin_earnhft_2023}.

The traditional quantitative research paradigm is fraught with several limitations. First, it adheres to a comprehensive pipeline that includes data processing, factor mining, machine learning, portfolio optimization, and algorithmic trading. Each of these steps demands significant research resources, including intensive labor and substantial time to identify effective "alphas." Furthermore, the optimization objectives across these pipeline stages often lack consistency, leading to suboptimal outcomes for the final trading strategy. Additionally, traditional task-specific quantitative modeling relies heavily on pre-defined scenarios, strategy tasks, and associated data, making it difficult to transfer these models directly to other strategy tasks. This reliance on "local" data not only limits the model’s potential but also exacerbates research costs, as quants are compelled to develop a distinct model for each strategy.

\begin{figure*}[ht]
	\centering
		\includegraphics[scale=0.4]{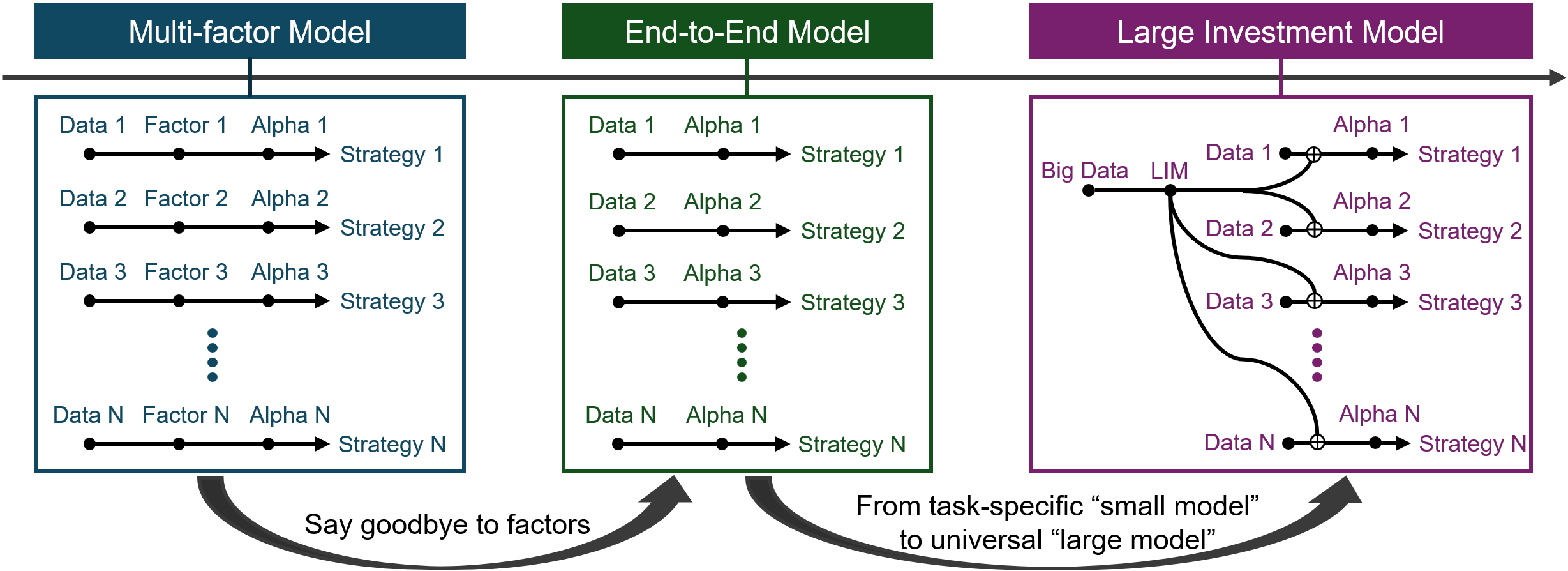}
	\caption{Three quant research paradigms: multi-factor model, end-to-end model and large investment model. }
        \label{fig_3paradigms}
    \vspace{-10pt}
\end{figure*}

In recent years, the rapid advancements in artificial general intelligence (AGI) have provided a unique opportunity to transform the quantitative research paradigm. Specifically, we discuss the shift toward a new modeling paradigm aimed at enhancing the efficiency and effectiveness of quantitative finance research. First, there is a clear transition from traditional multifactor modeling to state-of-the-art end-to-end modeling. Unlike multifactor modeling, which builds trading strategies incrementally through a research pipeline, end-to-end modeling seeks to directly generate the final trading strategy, bypassing intermediate steps such as factor mining, and producing predicted alphas, optimal positions, or even algorithmic trading orders. This approach has the potential to eliminate the labor-intensive factor mining process and significantly enhance the efficiency of quantitative research. Second, the shift from traditional task-specific modeling to universal modeling, akin to the "pretrained foundation model + fine-tuned task model" approach commonly used in large language models, is becoming increasingly prominent in quantitative investment. The foundation model, typically a universal model trained on a broad and diverse dataset (e.g., data spanning various countries, security markets, and trading assets), can be fine-tuned to optimize specific trading strategies. By combining the strengths of end-to-end modeling and universal modeling, we propose the Large Investment Model (LIM), a novel methodological framework for quantitative investment research. Figure~\ref{fig_3paradigms} illustrates the distinctions between multifactor modeling, end-to-end modeling, and universal modeling.


The remainder of this article is organized as follows. Section~\ref{sec_review_quant} provides a brief review of the data, strategies, and research pipeline in quantitative investment. Section~\ref{sec_lim} introduces the LIM framework and the underlying concepts. Details of the upstream foundation modeling and downstream strategy modeling within LIM are presented in Sections~\ref{sec_upstream} and~\ref{sec_downstream}, respectively. Section~\ref{sec_lim_system} discusses the architecture design for automated strategy generation and trading using LIM. Section~\ref{sec_research_directions} proposes several new research directions, and Section~\ref{sec_conclusion} offers concluding remarks. Finally, the appendix in Section~\ref{sec_experiment} presents the numerical experiments conducted to demonstrate the performance of LIM.

\section{Review on Quantitative Investment}\label{sec_review_quant}
Quantitative Investment relies on automated strategies built on various data to trade different instruments such as stocks, futures, bonds and options. This section briefly introduces common quantitative investment strategies and diverse data used for building these strategies. In addition, we introduce the classic multi-factor modeling which is wide used in quantitative investment. 

\subsection{Quant Strategies}
\begin{figure*}[ht]
	\centering
		\includegraphics[scale=0.27]{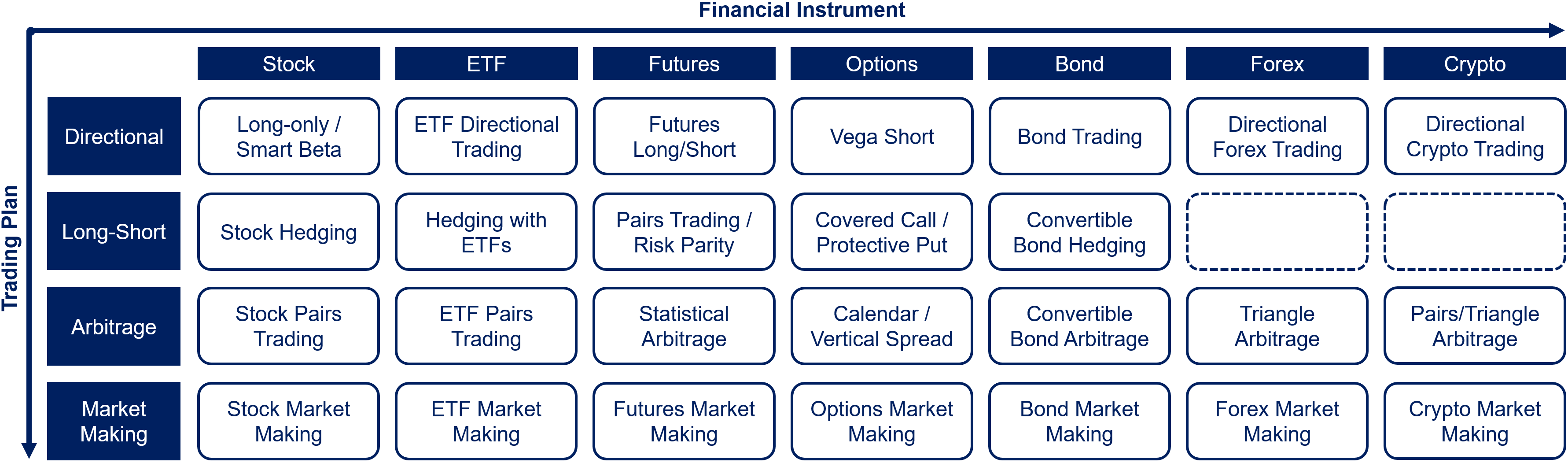}
	\caption{A strategy matrix for quantitative investment.}
    \label{fig_strategy_matrix}
    \vspace{-10pt}
\end{figure*}

A quantitative strategy is a systematic function or trading methodology used for trading financial instruments, such as stocks, options, and futures, in financial markets. These strategies are based on either predefined rules or trained models for making trading decisions and are typically the core intellectual property of a trading firm. A standard quantitative strategy should specify several configurations, such as the universe of financial instruments to be traded, the average holding period, and the trading frequency. Additionally, it should define the type of strategies employed. Figure \ref{fig_strategy_matrix} presents a strategy matrix that illustrates many popular trading strategy examples. The horizontal axis introduces a variety of financial instruments, including stocks, ETFs, futures, options, bonds, foreign exchange, and cryptocurrencies. The vertical axis contains four types of common trading approaches, each representing standard operations for trading various financial instruments and forming different strategies.

\begin{itemize}[noitemsep,topsep=0pt] 
    \item \underline{Directional trading} is a strategy used in financial markets that involves taking a position based on the anticipated direction of a security's price movement and profiting from these price changes by buying or selling securities accordingly. Popular directional trading strategies include trend following (identifying and following the direction of an existing trend, taking long positions in uptrends and short positions in downtrends), breakout trading (taking positions when the price decisively moves beyond support or resistance levels, with long positions on bullish breakouts and short positions on bearish breakouts), and contrarian trading (trading against the market trend by identifying overbought or oversold conditions and taking reverse positions).
    
    \item \underline{Long-short trading}, commonly used in hedge funds, is an investment strategy that involves taking both long and short positions in different securities to exclude market volatility effects (the ``Beta'' return) from the overall return and to profit from the ``Alpha'' return. A popular example is stock long-short selection, which predicts the ``best'' stocks to buy (long) and the ``worst'' stocks to sell (short) at each cross-section over time.  
    
    \item \underline{Arbitrage trading} is a strategy that exploits price discrepancies between different markets or financial instruments to achieve risk-free profits. Common arbitrage strategies include cross-exchange arbitrage (profiting from price discrepancies of a security across various exchanges), triangle arbitrage (typically used in forex and cryptocurrency markets to exploit exchange rate discrepancies by trading three different currencies), calendar spread arbitrage (for futures), convertible arbitrage (taking a long position in a convertible bond while shorting its underlying stock), and statistical arbitrage (trading pairs of correlated securities by longing the underpriced one and shorting the overpriced one).
    
    \item \underline{Market making trading} is a family of high-frequency strategies that provide liquidity to financial markets by continuously quoting both buy (bid) and sell (ask) prices for financial instruments, aiming to profit from the spread between these prices. For instance, a market maker might quote a bid price of \$100 and an ask price of \$100.10 for a stock. The market maker buys shares from a trader willing to sell at \$100 and sells them to another trader willing to buy at \$100.10, thereby profiting from the spread.
\end{itemize}

Additionally, trading frequency defines the duration for which assets are held in a portfolio and how often trades are executed. High-frequency trading typically involves holding positions for a few minutes or seconds, whereas low-frequency trading may involve holding assets for several months or years. The significant difference in holding periods between high-frequency and low-frequency trading leads to distinct considerations in strategy design. For example, asset capacity limitations and trading costs are critical issues in high-frequency trading, while managing drawdown risk is a primary concern in low-frequency trading.

\subsection{Data Diversity in Quant Modeling}
\begin{figure}[ht]
	\centering
		\includegraphics[scale=0.5]{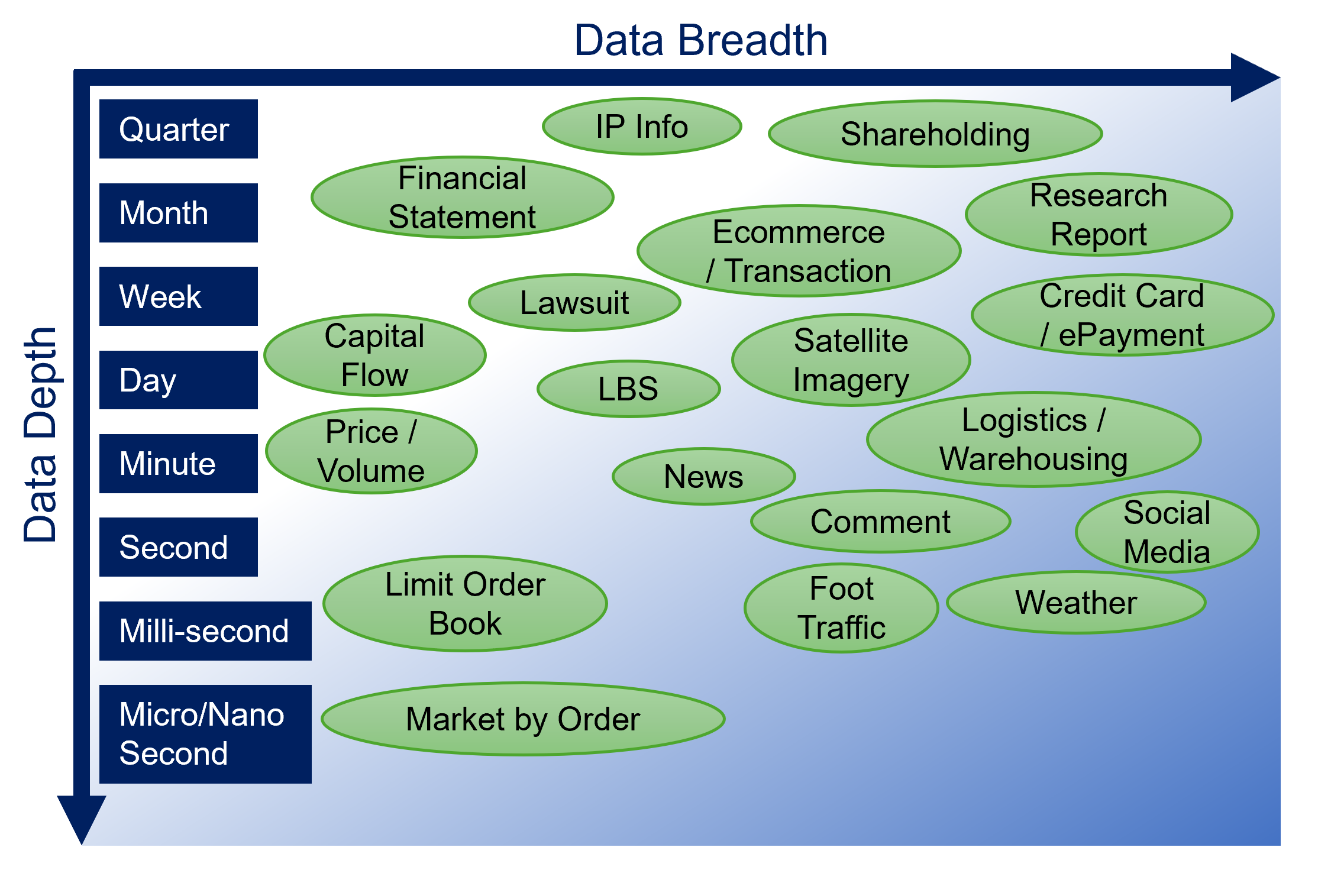}
	\caption{Spectrum of various financial data along depth and breadth.}
        \label{fig_depth_breadth}
    \vspace{-10pt}
\end{figure}

Modern quantitative investment harnesses a diverse array of data to develop statistical and machine learning strategies aimed at profitable trading. Figure~\ref{fig_depth_breadth} categorizes various types of financial data along two orthogonal dimensions: data depth and data breadth. Data depth refers to the granularity of data, which can span from several years at the macro level to mere nanoseconds at the micro level. Data breadth, on the other hand, indicates the diversity of the data, encompassing quote data (such as price/volume, limit order book (LOB) data, and market order flow), fundamental data (including financial statements, investment research reports, company announcements, and analysts' opinions), and a broad range of alternative data (such as e-commerce transactions, credit card/e-payment transactions, news and social media comments, satellite imagery, foot traffic, and supply chain data).

Different investment strategies rely on different types of financial data. For instance, high-frequency market-making strategies \cite{carmona2012high} focus on data depth by modeling granular LOB data to predict price movements over very short horizons. Horizontal spread arbitrage strategies \cite{Kou_calendarspread_2013} utilize minute-bar quote data, seeking to capitalize on price discrepancies in derivatives contracts (such as options or futures) with varying expiration dates. Stock technical trading strategies \cite{Han_technicalanalysis_2021} employ candlestick data (open, high, low, close prices) and trading volumes to generate buy/sell signals. Meanwhile, stock fundamental investing strategies \cite{WAFI2015939} analyze financial statements, analyst reports, and news data to evaluate the fundamental health and intrinsic value of public companies.

The rapid growth of internet and mobile technologies over the past decade has led to an explosion in the accumulation of big data. Financial institutions are increasingly integrating alternative data, such as credit card transaction data, web traffic data, and geolocation data, into their fundamental analysis and value investing strategies \cite{CAO2024102307,OLIVIER_alternativedata_2024}. This shift has significantly expanded the data breadth available for quantitative investment, enabling more comprehensive and nuanced analyses.

\subsection{Multifactor Quant Modeling}  
\begin{figure*}
	\centering
		\includegraphics[scale=0.38]{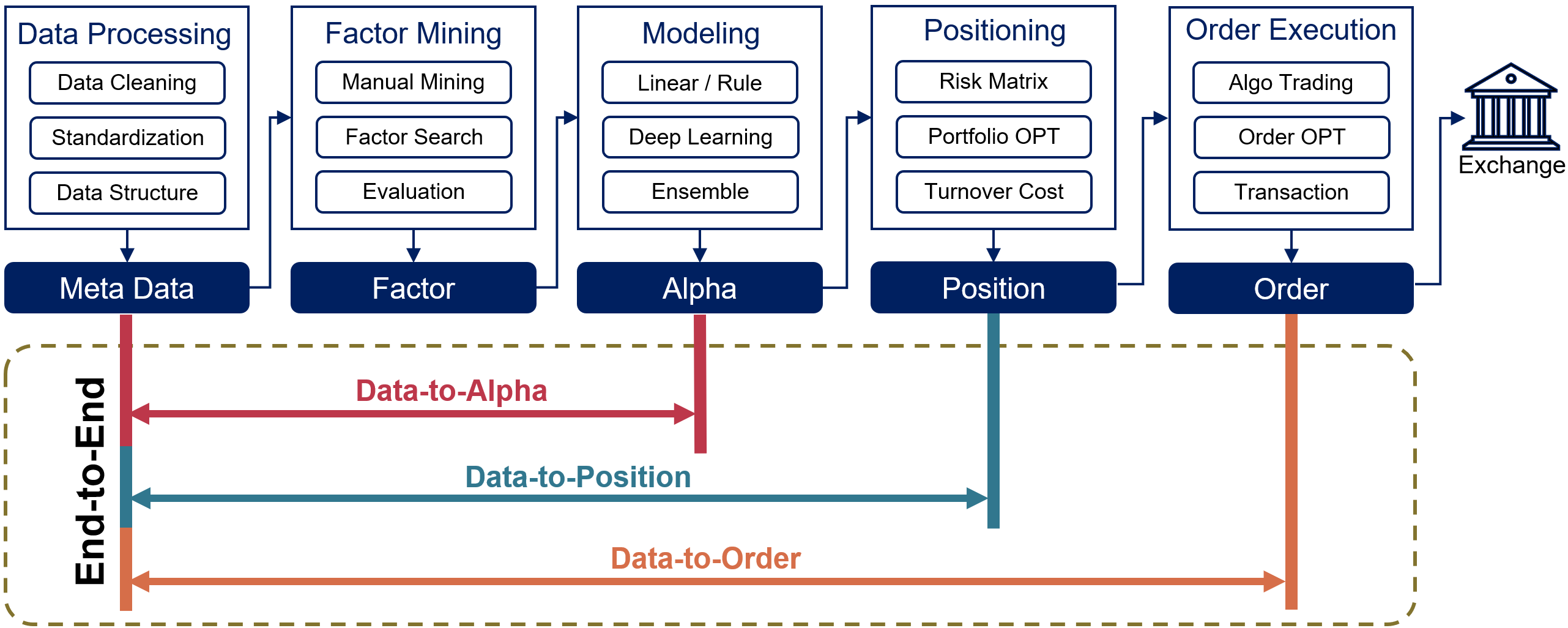}
	\caption{Quant research pipeline and various types of end-to-end quant modeling.}
        \label{fig_end2end_pipeline}
    \vspace{-10pt}
\end{figure*}

The quantitative research pipeline comprises several critical stages, including data processing, factor mining, alpha modeling, portfolio position optimization, and order execution optimization, as depicted in the blue section of Figure~\ref{fig_end2end_pipeline}. Among these stages, factor mining is particularly vital, as the quality of the factors significantly influences the performance of the alpha model, which subsequently impacts the overall returns of the final portfolio.

Factors are typically mathematical formulas or functions that capture signals predictive of trends in various financial instruments, such as stocks, futures, and foreign exchange. These factors can be derived from a wide range of data sources, including: (1) financial quote data, such as price, volume, and limit order book (LOB) information; (2) fundamental data, such as financial statements and research reports; and (3) alternative data sources \cite{Dixon_alternativedata_2022}, including credit card transactions, social media commentary, product reviews, satellite imagery, geolocation and climate data, shipping trackers, mobile app usage data, and news feeds. 

Traditionally, trading factors have been manually designed and constructed, relying heavily on market observations and the expertise of traders. However, there has been a growing shift toward the use of automatic factor mining techniques, such as genetic programming \cite{Chen_GP_factor_2021}, to improve the efficiency of factor construction and selection. Whether manually crafted or algorithmically discovered, these factors undergo rigorous back-testing. Only those factors that are both effective (``good'') and non-correlated (``diverse'') are retained and stored in the database for use in alpha modeling.

\section{Large Investment Models}\label{sec_lim}
The rapid advancement of artificial general intelligence (AGI) \cite{yin2024survey} in recent years has prompted a re-evaluation of the future trajectory of quantitative investment technology. Two pivotal aspects of AGI are end-to-end modeling and universal modeling. Notably, the success of GPT-like models \cite{brown2020language,openai2024gpt4} has highlighted the extraordinary potential of self-supervised generative learning \cite{balestriero2023cookbook}, which is grounded in a universal ``next-token prediction'' task. In this paradigm, a general-purpose foundation model is pre-trained on extensive datasets, and through appropriate fine-tuning \cite{han2024parameterefficient} for specific tasks, the model's predictive power is significantly enhanced, often surpassing models that were specialized from the outset for those tasks. 

This success invites an intriguing question: could the ``pre-training + fine-tuning'' paradigm be effectively applied to quantitative strategy research? If a robust foundational model \cite{zhou2023comprehensive} for quantitative investment can discover transferable trading patterns and investment logic across various instruments and markets, then quantitative strategy research might be reconceived as a fine-tuning task tailored to specific strategy requirements and investment scenarios. Such a paradigm shift could dramatically increase research efficiency in the field.

\subsection{End-to-End Modeling for LIM}
Since factors are essentially ``features'' that characterize instruments, their information is entirely derived from the original data. A natural question arises: can we build a predictive model without explicitly creating factors? The advent of deep learning and the end-to-end training paradigm presents a plausible technical route. End-to-end training refers to a modeling approach that directly learns the complex function linking raw inputs to final outputs, encompassing all intermediate stages. Figure~\ref{fig_end2end_pipeline} illustrates three types of end-to-end modeling, each starting from the original meta-data (raw data with standardized and simple preprocessing) and leading to different outputs: alpha predictions (e.g., predicted returns over a future horizon), portfolio positions (e.g., the optimal position size at the next trading point), or trade orders (e.g., the optimal order size in the next second for trading).

Consider the example of data-to-position modeling, where the tasks of alpha generation and optimal position determination are integrated within a single deep learning model. There are several methods available for directly outputting position sizes. One such method involves calculating Markowitz-optimal position sizes \cite{markowitz1952portfolio} based on ground-truth returns and using these computed values as labels for training the deep learning model. Another approach frames the reinforcement learning reward function in terms of the portfolio's Sharpe ratio, which is derived from the predicted position sizes. This subsection explores a third technical approach: designing a novel loss function for deep learning, based on the predicted portfolio Sharpe ratio, to directly determine the optimal position sizes. This method enables the deep learning model to learn the optimal portfolio allocation by optimizing a loss function that directly reflects the trade-off between return and risk, as quantified by the Sharpe ratio. Such an approach has the potential to improve portfolio performance by allowing the model to focus on maximizing the risk-adjusted return throughout the training process. Mathematically, suppose we have a universe of $m$ stocks, each with a record of $n$ time points. Let $r_{t,s}$ and $w_{t,s}$ ($1 \le t \le T$ and $1 \le s \le S$) denote the return and position size of stock $s$ at time point $t$, respectively. The portfolio return at time point $t$ is $r_t = \sum_{s=1}^S w_{t,s} r_{t,s}$. Let $\V{r} = \{r_t\}_{t=1}^n$ represent a sequence of $n$ portfolio returns over time, then the negative logarithmic Sharpe ratio of the portfolio can be defined as:
\begin{equation}
    \F{L}(\V{X} | \V{w}, \V{\theta}) = -\F{log}(\F{E}(r)) + \F{log}(\F{\sigma}(r))
\end{equation}
This serves as the loss for the end-to-end deep learning quant model. Note that $r$ depends on $r_{t,s}$, which is a function of the input data sequence $\V{X} = \{x_t\}_{t=1}^n$, position $\V{w}=\{w_1, w_2,\ldots, w_m\}$ ($m$ is the number of stocks in the universe), and deep neural network parameters $\V{\theta}$. Since $\F{L}(\V{X} | \V{w}, \V{\theta})$ is a differentiable function of $\V{w}$ and $\V{\theta}$, we can build an end-to-end neural network with this loss function to output the optimal portfolio positions corresponding to an optimal Sharpe ratio on the training data.

Recent literature has seen a growing interest in this area, with notable contributions such as Deep Inception Networks (DINs)~\cite{liu_deep_2023} and End-to-End Active Investment (E2EAI)\cite{wei_e2eai_2023}. DINs introduce end-to-end systematic trading strategies by extracting time-series and cross-sectional features directly from daily price returns, outputting position sizes by optimizing the Sharpe ratio of the entire portfolio during training. Similarly, E2EAI presents an end-to-end neural network model that spans the entire quantitative research process, from factor selection and combination to stock selection and portfolio construction. Both approaches bypass traditional factor mining and alpha prediction steps, directly outputting optimal positions and enabling traders to compute the difference between new and old positions to determine subsequent orders. Another significant contribution is DeepLOB~\cite{zhang_deeplob_2019}, which constructs a large-scale deep learning model to predict price movements directly from limit order book (LOB) data of cash equities. Notably, the authors emphasize that DeepLOB generalizes well to instruments not included in the training set, demonstrating the model's ability to extract universal features. The team further extends deep learning models to more granular micro-structure data in~\cite{zhang_deep_2021}, concluding that an ensemble of MBO (market by order) and LOB data enhances forecasting accuracy. Unlike the straightforward end-to-end modeling in~\cite{zhang_deeplob_2019} and~\cite{zhang_deep_2021}, the work in~\cite{jiao_microstructure-empowered_2023} proposes an upstream pretraining framework to extract alphas from order flow data, applicable across various granularities and scenarios. This approach also inspires the large investment model proposed in this paper.

Compared to traditional quant research pipeline (blue part of Figure~\ref{fig_end2end_pipeline}), end-to-end modeling offers several advantages: 1) in traditional quantitative research, the optimization goals of individual modules are usually inconsistent. For instance, each factor is evaluated and selected based on criteria that primarily concern the factor itself, rather than its interaction with other factors. As a result, a ``good'' factor with a high information coefficient (IC) \cite{zhang2020information} or Sharpe ratio \cite{Hayette_sharpe} may negatively impact an alpha model due to complex interactions with other factors, while a ``bad'' factor might significantly contribute to the model. 2) The formulaic nature and operator space of factors can limit their representational capability. Almost all operators (e.g., $rank(\cdot)$, $ts\_{max}(\cdot)$, etc.) that define formulaic factors are simple algebraic functions, and the representational power of their combinations is difficult to compare with deep neural networks. Therefore, with sufficient sample size, end-to-end modeling has a higher ceiling than traditional multi-factor modeling. 3) Factor mining is a labor-intensive and time-consuming process, especially for building and selecting factors by hand. On the other hand, end-to-end modeling throws these ``dirty work'' to deep learning algorithms and may reduce the cost significantly. 

Compared to the traditional quantitative research pipeline (illustrated in the blue section of Figure~\ref{fig_end2end_pipeline}), end-to-end modeling presents several significant advantages. First, in the traditional approach, the optimization goals of individual modules often lack consistency. For instance, factors are typically evaluated and selected based on criteria that focus primarily on the factor itself, rather than on its interaction with other factors. This can lead to situations where a ``good'' factor, characterized by a high information coefficient (IC) \cite{zhang2020information} or Sharpe ratio \cite{Hayette_sharpe}, may inadvertently have a detrimental effect on an alpha model due to complex interactions with other factors. Conversely, a ``bad'' factor could unexpectedly enhance model performance. Second, the formulaic nature and operator space of traditional factors limit their representational capacity. Most operators (e.g., $rank(\cdot)$, $ts\_{max}(\cdot)$) that define formulaic factors are simple algebraic functions. The representational power of these combinations is difficult to compare with that of deep neural networks. Thus, with a sufficiently large sample size, end-to-end modeling has a higher potential for performance than traditional multi-factor modeling. Third, factor mining is a labor-intensive and time-consuming process, particularly when factors are manually constructed and selected. In contrast, end-to-end modeling delegates these ``dirty tasks'' to deep learning algorithms, potentially reducing the overall cost and effort significantly.

\subsection{Universal Modeling for LIM}
\begin{figure*}[ht]
	\centering
		\includegraphics[scale=0.3]{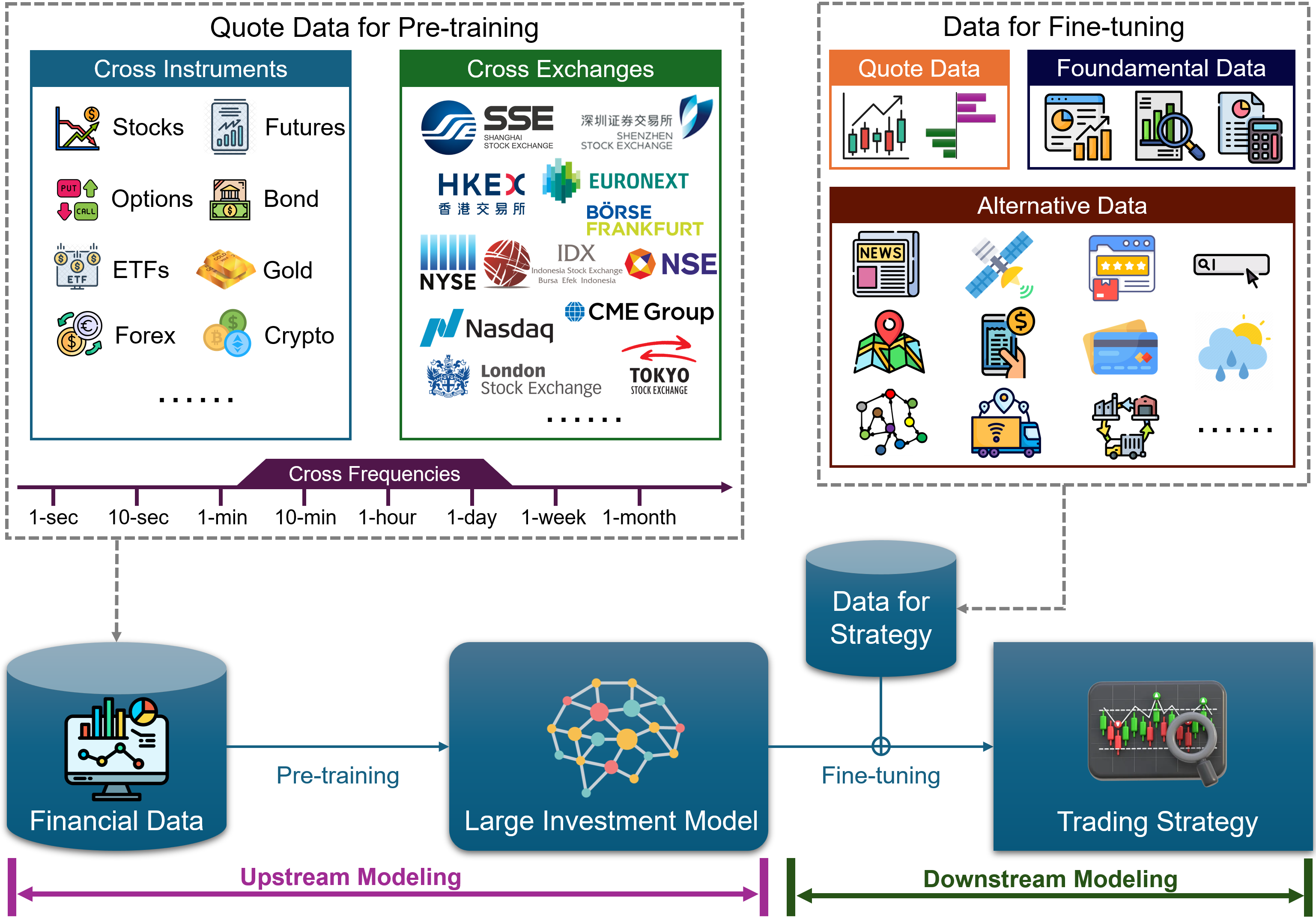}
	\caption{Workflow of large investment model.}
        \label{fig_LIM}
    \vspace{-10pt}
\end{figure*}
The universality of LIM should encompass at least the following three aspects:
\begin{enumerate}[noitemsep,topsep=0pt]
    \item \underline{Cross-instrument universality}. Given quote data, can a machine-mined pattern for predicting stock trends be applied to predict trends in futures or bonds? Logically, many trading patterns reflect traders' intentions and behaviors, and it is natural for some common patterns to be shared across various instruments. Empirically, many technical or price/volume factors are useful not only for stock prediction but also for bonds, futures, and even cryptocurrencies. This suggests the feasibility of training a general-purpose upstream model with data from various instruments and fine-tuning this model with specific data for each instrument.
    \item \underline{Cross-exchange universality}. Stock trading across different exchanges may exhibit common patterns or signals, especially for technical indicators or strategies based on quote data (and sometimes news data). This observation motivates the development of a ``universal'' model using data from multiple exchanges, which can then be applied to trade equities in specific markets. Similarly, many other instruments (e.g., futures, bonds) share cross-exchange patterns, making them suitable for pretraining the foundational quant model.
    \item \underline{Cross-frequency universality}. Patterns often persist across different data frequencies, such as 1-second, 15-second, 1-minute, 20-minute, 1-hour, and daily candlesticks. Training on data from the same instrument across various frequencies can significantly enhance the sample size, which is crucial for improving the performance of deep learning models.
\end{enumerate}

As shown in Figure~\ref{fig_LIM}, the architecture follows a typical pretraining-finetuning structure. The upstream foundation model is a self-supervised generative model pretrained on diverse data, acquiring financial analysis, prediction capabilities, and data generation skills through exposure to a wide array of data, corpora, and knowledge. The downstream process involves building specific quantitative investment strategies by fine-tuning the upstream model with task-specific data and outputting a prediction model for the quantitative trading strategy. Analogous to GPT for natural language processing, LIM aims to serve as an artificial general intelligence system for quantitative investment. First, the upstream model acts as a generative foundation for quantitative finance, simplifying strategy formulation into a unified framework akin to a ``next-token prediction'' problem \cite{li2024mechanics}. Utilizing self-supervised learning, the model efficiently learns representations from various financial data across different exchanges and instruments, capturing nuanced market patterns and relationships. This predictive approach streamlines strategy development by transforming traditional task-specific modeling into a more generalized sequence prediction problem. Consequently, the model can infer future market conditions based on historical data, similar to how language models predict the next word in a sentence, thereby unlocking new potential for quant strategy research, algorithmic trading, risk assessment, and portfolio management in a more automated and scalable manner. To maximize the universality of the upstream model, we define it as a single-instrument processor. Therefore, strategies dependent on multiple trading instruments (such as pairwise arbitrage \cite{Christopher_pairstrading_review_2017} or cross-sectional stock alpha strategies \cite{1bbb43a3-4de0-3ecb-be89-4bff15e655aa}) will be modeled during the downstream fine-tuning phase. Second, the downstream model fine-tunes the upstream model according to specific task requirements to develop quantitative trading strategies, including momentum strategies, mean-reversion strategies, pairs trading strategies, triangular arbitrage strategies, calendar spread arbitrage, and cross-sectional hedging strategies. Given the diverse specifications and configurations of different tasks, downstream modeling employs a range of approaches. In addition to investment strategies, downstream tasks can also include risk models or stochastic simulators (see Section~\ref{sec_research_directions} for details) to simulate various market scenarios.

\section{Upstream Foundation Model}\label{sec_upstream}
\begin{figure*}[ht]
    \centering
    \includegraphics[scale=0.4]{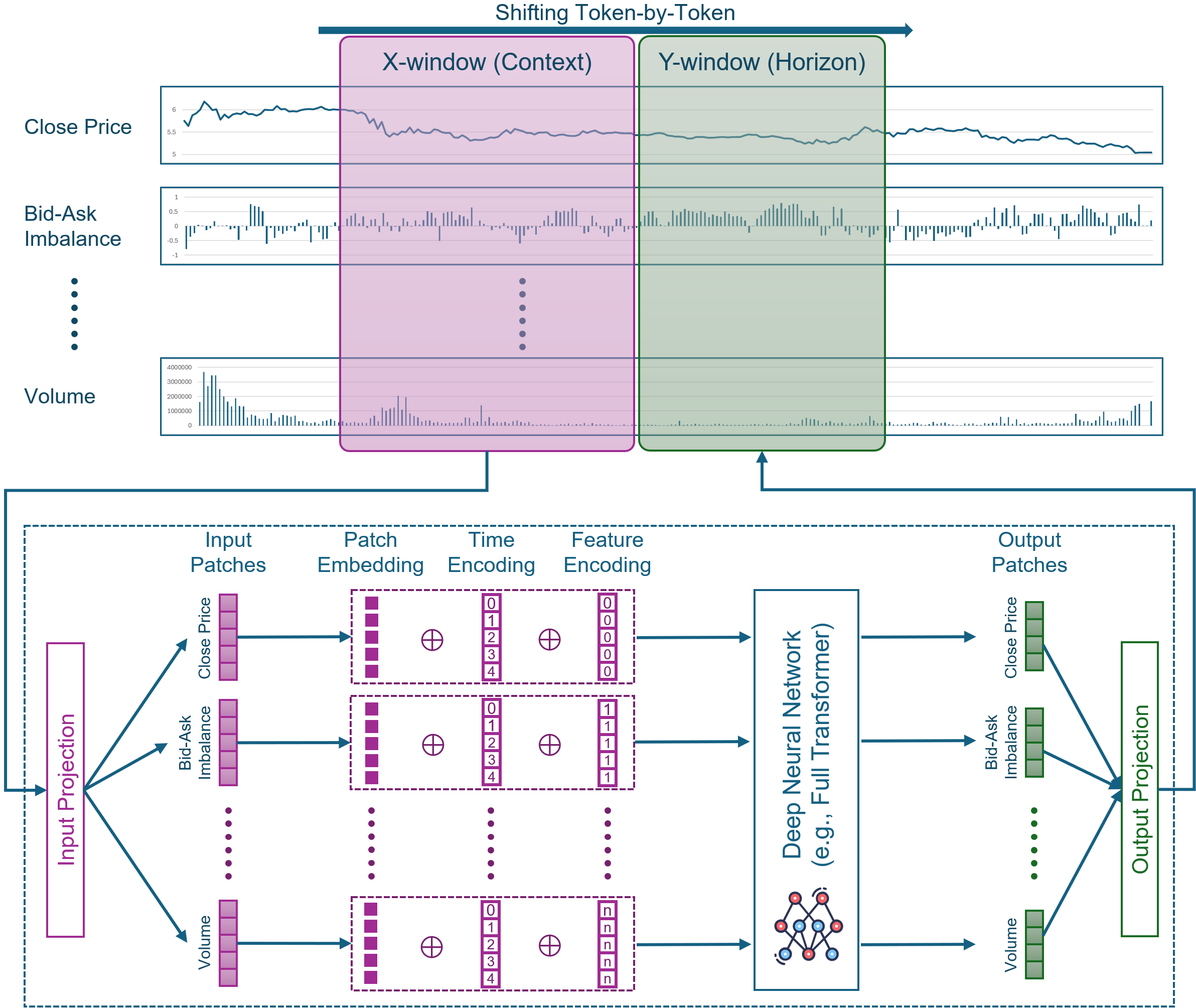}
    \caption{Illustration of the workflow for the upstream foundation model. Financial quote data (price/volume and LOB) are used to construct the backbone of the foundation model, while other data types, including fundamental and alternative data, are integrated into the backbone aligned with the time axis and investment instruments.}
    \label{fig_LIM_detail}
    \vspace{-10pt}
\end{figure*}
As illustrated in Figure~\ref{fig_LIM_detail}, the upstream modeling focuses on developing a universal foundation model for quantitative investment. The goal of the upstream pretraining foundation model is to be as general as possible, enabling it to address a broad spectrum of financial time-series prediction problems.

\subsection{Problem Formulation}
We formulate the foundation model in the upstream pre-training stage as follows. Suppose we have a dataset of $M$ multivariate time series $\mathcal{D} = \{\mathbf{X}^{(m)}\}_{m=1}^M$, where each multivariate time series $\mathbf{X}^{(m)} \in \mathbb{R}^{T^{(m)} \times p}$ has a length $T^{(m)}$ and dimension $p$. For each time point $t$ ($1 \le t \le T$), we define two sliding windows: a look-back context window of length $L_X$ that defines the input $\mathbf{X}_{t-L_X+1:t}$, and a look-forward horizon window of length $L_Y$ that defines the output $\mathbf{X}_{t+1:t+L_Y}$. The foundation model is a function $f: \mathbb{R}^{L_X \times p} \rightarrow \mathbb{R}^{L_Y \times p}$. Given parameters $\mathbf{\Theta}$, we aim to satisfy the relationship $\mathbf{X}_{t+1:t+L_Y} \approx f(\mathbf{X}_{t-L_X+1:t} \mid \mathbf{\Theta})$. 

To estimate the mapping function $f$, various deep learning models can be utilized by minimizing the loss function:
\begin{equation}
    \frac{1}{M} \sum_{m=1}^M \frac{1}{T^{(m)}} \sum_{t=1}^{T^{(m)}} L(\mathbf{X}_{t+1:t+L_Y}, f(\mathbf{X}_{t-L_X+1:t} \mid \mathbf{\Theta}))
\end{equation}
where $L(\cdot,\cdot)$ is a predefined distance metric, such as mean-square error (MSE) or cross-entropy loss~\cite{10.5555/3618408.3619400}. Under the assumptions of continuity and differentiability, this loss function can be differentiated to derive the gradient, which is used to optimize the parameters during the training of deep neural networks.

\subsection{Modeling Principle}
To build a foundation model for quantitative investment, several conditions must be carefully considered:
\begin{enumerate}[noitemsep,topsep=0pt] 
    \item \underline{Single-instrument time series}: To accommodate a wide range of trading strategies and investment scenarios, the upstream model should focus on single financial instrument time series (note that the time series may be multivariate for a single financial instrument). Modeling involving multiple financial instruments (e.g., cross-sectional stock selection for the S\&P 500) should be reserved for downstream tasks.
    \item \underline{Trading cost and rules}: Significant differences in trading rules and transaction costs across exchanges exist. For example, some stock exchanges operate on a T+1 trading basis, where stocks bought today cannot be sold before the next trading day, while others use a T+0 system, allowing for same-day trading. These differences affect the timing of cash flows, market patterns, and overall strategy design and execution. Since the LIM foundation model is expected to learn common market patterns, these differences across exchanges are ignored, treating the problem as a pure time-series prediction task.
    \item \underline{Patching and masking}: A time-series patch serves as an analogue to a token in language models, and patching has been shown to improve performance. This approach also enhances inference speed by reducing the number of tokens fed into the transformer by a factor equivalent to the patch length. Additionally, employing a random masking strategy can induce adaptive window lengths, allowing the model to experience all possible context lengths, ranging from 1 to the maximum context length, ensuring robust performance across various context lengths.
    \item \underline{Data choice}: Financial data for investment can be broadly categorized as quote data, fundamental data, and alternative data. Among these, quote data at various frequencies are used to pre-train the foundation model because they are regularly sampled and can be collected across different exchanges. Fundamental and alternative data are used in downstream tasks to build strategy models.
\end{enumerate}

\subsection{Design of Foundation Model}
Figure~\ref{fig_LIM_detail} illustrates the construction of an upstream foundation model for LIM. This model is designed to predict future tokens within a Y-window (horizon) based on data from an X-window (context) that covers a fixed historical period. The X-window data are fed into the modeling module, serving as the input for the backbone deep learning model. This model is trained on financial quote time-series data, incorporating various variables (meta-features) such as closing price, bid-ask imbalance, returns, and trading volume, with the aim of predicting the same set of variables within the Y-window. To enhance computational efficiency, time-series segmentation techniques such as patching~\cite{nie2023timeseriesworth64} are applied within the windows. Additionally, patch masking strategies~\cite{das2024decoderonlyfoundationmodeltimeseries} are employed to improve the quality of self-supervised learning and increase the flexibility of window length during model training.

An example architecture of the deep neural network is illustrated in Figure~\ref{fig_LIM_detail}. For each variable in the time series, the input data from the X-window are first transformed into input patch vectors via an input projection, which then generates patch embedding vectors. These embeddings are concatenated with a time encoding vector to capture temporal information and a feature encoding vector to identify the specific variable being used. After processing through a deep neural network (e.g., a Transformer), the model outputs patches corresponding to each variable. These outputs are then merged and converted back to the original time-series granularity through an output projection, ultimately generating predictions for the variables in the Y-window.

\section{Downstream Task Model}\label{sec_downstream}
The downstream workflow bridges the foundation model from the upstream process with the final strategy development task. Unlike the foundation model, which primarily relies on quote data, downstream modeling can incorporate a wide variety of task-specific data sources, including news, supply chain information, satellite imagery, earnings call transcripts and so on. These diverse data types can be categorized into graph data, textual data, image data, numerical data, audio data, and video data. To effectively utilize these additional inputs, we employ specialized embedding techniques tailored to each data type, enabling the model to integrate and leverage the unique information contained within these varied structures.

\subsection{Data Alignment and Standardization}
\begin{figure*}[h]
	\centering
		\includegraphics[scale=0.4]{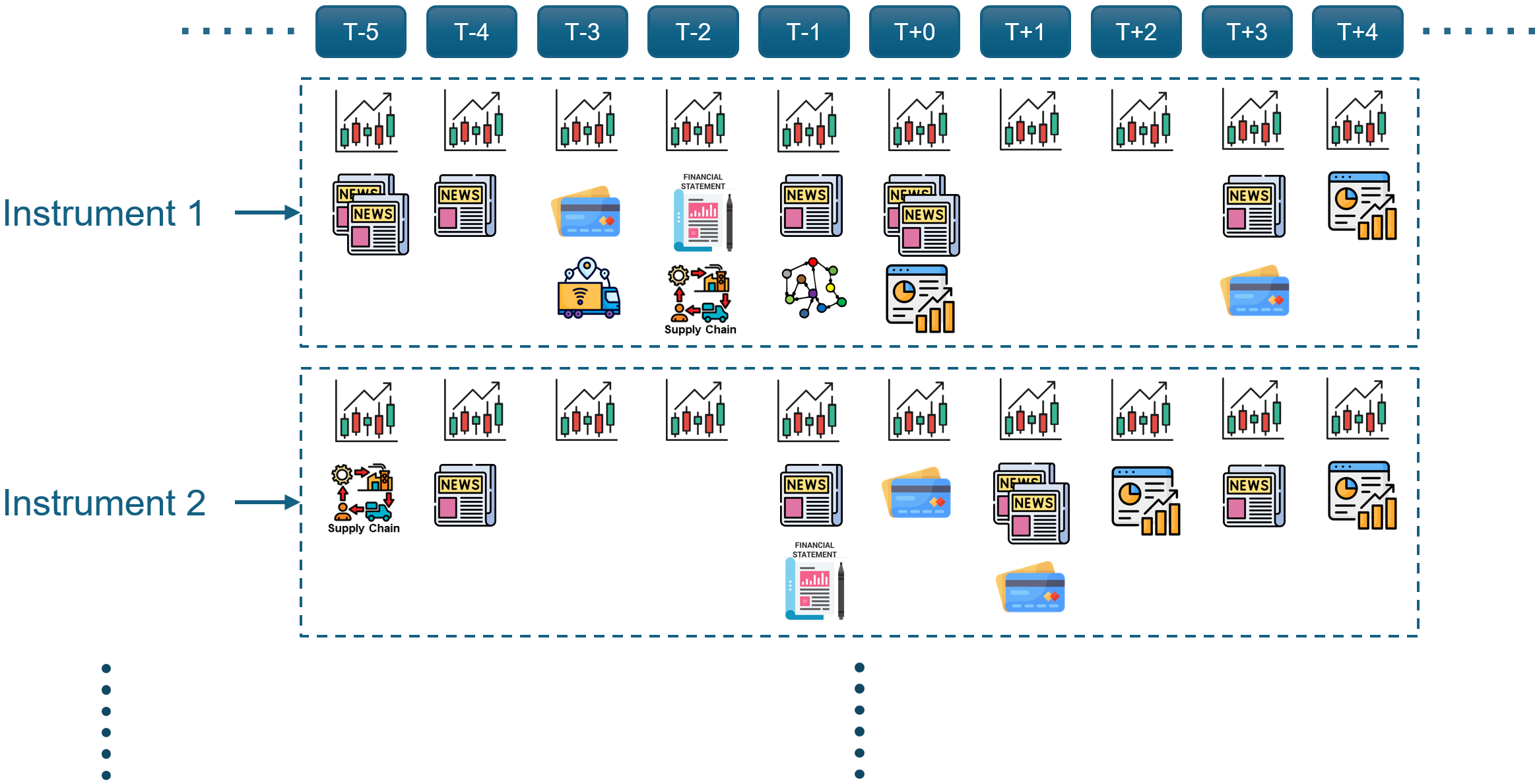}
	\caption{Align diverse data with time patch and financial entities.}
        \label{fig_alignment}
    \vspace{-10pt}
\end{figure*}
Handling diverse data types is crucial for developing robust quant models. Fundamental data, such as financial statements, and alternative data, such as social media sentiment and satellite imagery, often exhibit irregularities in their time series. This irregularity poses significant challenges for data preprocessing, necessitating a methodical approach to align data accurately with corresponding time points and financial instruments. 

Proper alignment is essential to ensure that the input data used for model training and evaluation is coherent, consistent, and reflective of true market conditions. First, a fundamental step in data preprocessing is temporal alignment. Since fundamental and alternative data points do not always coincide with regular intervals, it's necessary to synchronize these data points to a common timeline that matches the trading dates or specific events relevant to the investment strategy. Techniques such as interpolation or nearest-neighbor methods can be employed to estimate missing values and align the data with the appropriate time stamps. This alignment ensures that the models receive continuous and accurate inputs, thereby enhancing their predictive accuracy and reliability. Second, instrument alignment is equally critical in the preprocessing phase. Given that different financial instruments (e.g., stocks, bonds, derivatives) may have unique characteristics and response patterns to various data inputs, aligning data to the correct instrument is imperative. This involves mapping the fundamental and alternative data to the specific instruments they relate to, ensuring that each data point is correctly attributed. For instance, corporate earnings reports must be matched to the corresponding company's stock, while industry-wide metrics should be appropriately linked to all relevant securities within that industry. This precise mapping enhances the granularity and relevance of the data, allowing for more targeted and effective modeling.

Furthermore, aligning data across time points and instruments also involves normalization and standardization processes. These processes adjust data to a common scale, which is crucial when combining disparate data sources. Normalization mitigates the impact of outliers and scales differences, while standardization ensures that all data inputs have a consistent distribution, facilitating more efficient training of machine learning models. By incorporating these preprocessing steps, the data fed into AI models becomes more robust, reducing noise and improving the overall quality of the predictions.

\subsection{Model Fine-tuning}
Fine-tuning the LIM foundation model, which is initially pre-trained with quote time series, is a critical step in enhancing its applicability to broader quantitative investment strategies. The foundation model, built primarily on historical price data, needs to be adapted to incorporate additional layers of information to better understand market dynamics. By integrating more fundamental and alternative data alongside the quote data, the fine-tuning process enriches the model with a comprehensive dataset, making it more adept at predicting market movements and refining investment strategies. The use of advanced fine-tuning methods and optimization techniques ensures that the model not only retains its foundational strengths but also evolves to meet current market demands. These enhancements enable the development of more accurate, reliable, and sophisticated investment strategies, ultimately contributing to improved performance in quantitative investment.

The fine-tuning process involves adding various new data types, including numerical data such as financial ratios and technical indicators, graph data representing relationships between entities (e.g., corporate networks and supply chains), and unstructured data like images, videos, and audio. For instance, satellite imagery can offer insights into industrial activities, while social media sentiment analysis can provide contextual understanding of market sentiments. This diverse data integration allows the model to capture a wider array of market signals and develop more sophisticated strategies.

Adapting the foundation model to meet current strategy development demands involves employing several common fine-tuning methods appropriate for quantitative investment. Transfer learning~\cite{5288526} is a pivotal technique where the pre-trained model is fine-tuned on a new, specific dataset, allowing it to retain its learned knowledge while becoming more specialized. Feature-based Transfer Learning~\cite{daume-iii-2007-frustratingly} adjusts only the last few layers, leveraging the previously learned features, while Fine-tuning Entire Models~\cite{10.5555/2969033.2969197} retrains the entire network to adapt to the new domain. Layer-wise Fine-Tuning~\cite{howard-ruder-2018-universal} is a technique where different layers of the model are fine-tuned at different rates, typically starting from the last layers and progressively fine-tuning earlier layers. Knowledge Distillation~\cite{hinton2015distilling} is another approach, where a smaller model (student) is fine-tuned using the outputs of a larger pre-trained model (teacher) as soft targets. Parameter-Efficient Tuning methods, such as Low-Rank Adaptation (LoRA)~\cite{hu2021lora} and Adapter modules~\cite{pfeiffer2020adapterfusion}, introduce a small number of trainable parameters to existing pre-trained models, allowing for efficient fine-tuning with fewer resources. These algorithms are chosen based on the nature of the task, the available data, and computational resources, balancing the need for adaptation with the risk of overfitting.
\begin{figure*}[h]
	\centering
		\includegraphics[scale=0.38]{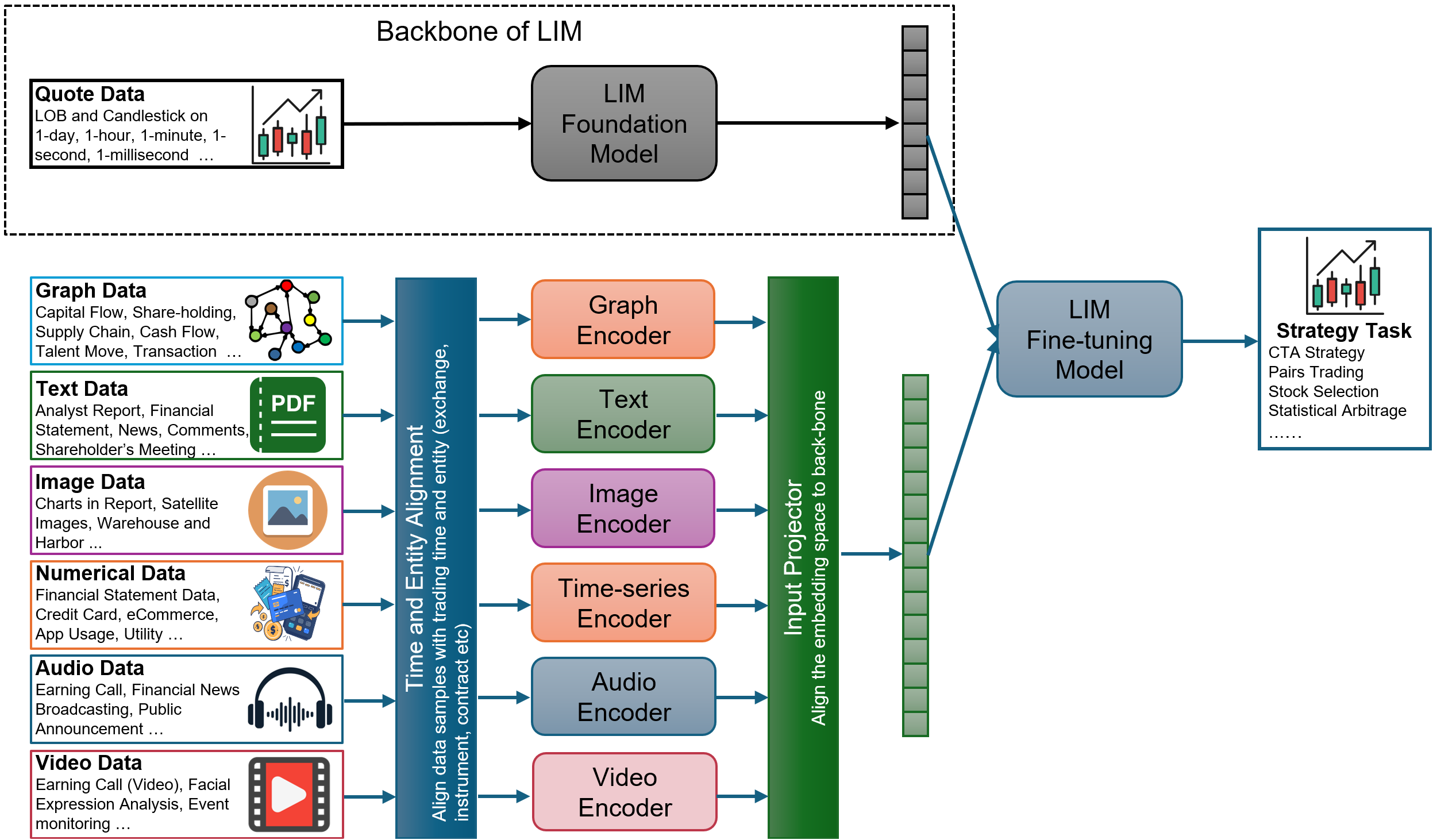}
	\caption{Downstream workflow for large investment model.}
        \label{fig_downstream}
    \vspace{-10pt}
\end{figure*}

\subsection{Various Types of Downstream Tasks}
Quantitative investment encompasses a broad range of strategy types. Given that the upstream foundation model primarily serves as a predictor for single time-series data, the downstream process becomes crucial for strategies that involve multiple instrument time-series. These include strategies such as cross-sectional stock selection, pairs trading, and various complex arbitrage approaches. Below, we outline several typical strategy scenarios and describe the corresponding downstream processing tasks:

\begin{itemize}[noitemsep,topsep=0pt] 
    \item \underline{Fundamental Investing}. Fundamental investing, which often involves low-frequency trading, relies heavily on fundamental data sources such as analysts' reports, financial statements, news articles, and other alternative data related to company performance and operations. In the downstream process, these data types are combined with the embeddings generated by the pretrained model to fine-tune a new prediction model focused on fundamental analysis. Due to the typically small sample size in low-frequency trading, the downstream model is often limited to predicting the next alpha signal. Subsequently, portfolio positions are optimized using a Markowitz optimizer, and order execution strategies are derived from algorithmic trading techniques.
    
    \item \underline{Statistical Arbitrage}. Statistical arbitrage~\cite{RePEc:zbw:iwqwdp:092015} strategies involve trading two or more historically correlated assets based on deviations from their mean or expected relationship. For instance, in pairs trading, when one asset outperforms its counterpart, the strategy may involve selling or shorting the overvalued asset while buying or going long on the undervalued asset, with the expectation that the spread will revert to its historical average. In this context, the upstream foundation model embeds the assets used in pairs trading into latent vectors. These vectors are then processed by the downstream model to predict optimal trading times.
    
    \item \underline{Lead-Lag Strategy}. The lead-lag strategy \cite{Li_leadlag_2022} involves trading two assets where one asset (the "leading" asset) is anticipated to influence the performance of the other (the "lagging" asset). Unlike pairs trading, where assets are traded in opposite directions, the lead-lag strategy involves trading only the lagging asset based on the trend of the leading asset. In this scenario, the downstream model is fine-tuned to take embeddings of both the lead and lag time series as input, and outputs predictions for the lagging series.
    
    \item \underline{Cross-Sectional Strategy}. Cross-sectional strategies~\cite{Engelberg_McLean_Pontiff_Ringgenberg_2023} differ from time-series approaches in that they involve trading a broad universe of assets simultaneously based on predicted alphas for the same time horizon. Common cross-sectional strategies include stock selection and long-short hedging. During the downstream process, the entire cross-section of assets is input into the fine-tuning model as a single sample. Multiple cross-sectional samples from different time points are used to train the downstream model, ultimately guiding the selection of stocks for buy/sell or long/short trades based on the predicted horizons.
\end{itemize}

\section{System Architecture for LIM}\label{sec_lim_system}
\begin{figure*}
	\centering
		\includegraphics[scale=0.55]{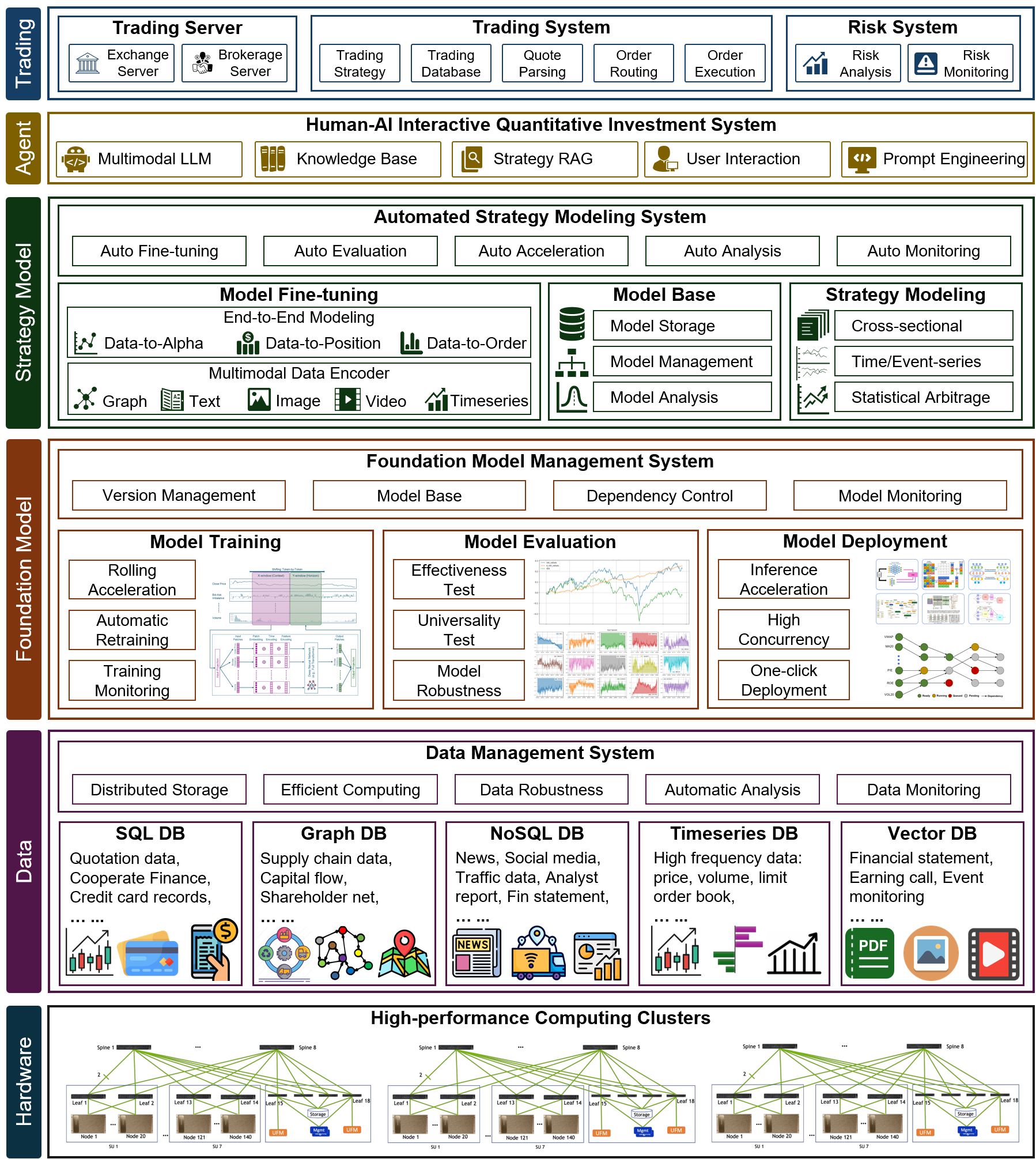}
	\caption{Architecture design for an LIM system platform. The platform is composed of hardware layer, data layer, foundation model layer, strategy model layer, agent layer and trading layer, and each layer consist a number of system modules.}
        \label{fig_LIM_system}
    \vspace{-10pt}
\end{figure*}
This section outlines the construction of a real-world system founded on the Large Investment Model (LIM) methodological framework. This comprehensive system supports the entire modeling pipeline, including computing infrastructure, data computation and storage, foundation modeling and management, automated strategy modeling, human-AI interaction agents, and a low-latency trading system.

\subsection{Computing and Data Infrastructure}
Building large-scale investment models requires the integration of high-performance computing (HPC) platforms to manage the complexity and volume of data. These platforms facilitate efficient training and execution of models, ensuring scalability to accommodate growing data volumes and computational demands \cite{reuter2019hpc}. The architecture of HPC systems can be specially optimized for financial time-series modeling, enhancing the performance of deep learning models applied to these data types.

An effective and reliable data system is crucial for deploying large investment models. This system must support a variety of database types to meet different data storage and retrieval needs:
\begin{itemize}[noitemsep,topsep=0pt] 
    \item SQL databases manage relational data such as candlestick data, financial statement records, and transaction data, ensuring robust data integrity and complex query capabilities \cite{stonebraker2010sql}.
    \item Graph databases are excellent for handling data with complex relationships and interconnections, useful for analyzing supply chain networks and stock relationships \cite{angles2008graph}.
    \item NoSQL databases are suited for unstructured and semi-structured data, like social media feeds and news articles, offering flexibility and scalability \cite{han2011nosql}.
    \item Time-series databases specialize in managing temporal data, crucial for tracking financial market data and economic indicators \cite{dunning2019time}.
    \item Vector databases store high-dimensional embedding vectors that characterize various data representations, managing metadata from diverse alternative data sources.
\end{itemize}

To complement this diverse data system, constructing a high-performance data computation system is essential for accelerating data preprocessing tasks. Utilizing distributed computing frameworks such as Apache Spark allows for parallel processing of large datasets, significantly reducing the time required for data cleaning, transformation, and integration \cite{zaharia2016apache}. In-memory computing technologies like Apache Ignite and Redis enhance processing speeds by storing data in RAM for quick access and manipulation \cite{ignite2019}. For real-time trading, stream processing frameworks like Apache Flink provide continuous ingestion and processing of real-time data streams, ensuring timely decision-making for investment models \cite{carbone2015apache}.

Building a highly reliable data system supports large-scale investment models by ensuring data integrity, availability, and security. Distributed storage solutions such as the Hadoop Distributed File System (HDFS) and Amazon S3 offer scalable and fault-tolerant storage, distributing data across multiple nodes to ensure redundancy and high availability \cite{shvachko2010hadoop}. Efficient computing is further achieved through algorithm optimization and advanced hardware, enhancing computational efficiency and enabling faster, more accurate predictions. Data robustness is maintained through rigorous validation and cleansing procedures, alongside redundant storage systems to mitigate the risks of data corruption and loss. Automated data analysis and comprehensive computation monitoring with tools like Prometheus and Grafana provide real-time insights into system performance, facilitating proactive management and optimization of computing resources \cite{turnbull2018monitoring}.

 \subsection{Systems for Foundation Model}
The core of an LIM system begins with the construction of a comprehensive module for foundation modeling training. This module integrates several advanced technical features to ensure efficiency and accuracy. A pivotal feature is rolling training acceleration, achieved through incremental learning or continuous learning techniques, allowing the model to continually update itself with new data without the need for retraining from scratch, significantly reducing both training times and computational resource demands \cite{polino2018modelcompression}. Automatic retraining is implemented to keep the model current by periodically integrating the latest data, thus maintaining the model's performance and relevance over time. Moreover, training monitoring is crucial for managing the training process, involving real-time tracking of metrics such as training loss and validation accuracy to facilitate the early detection of issues and enable timely adjustments \cite{goodfellow2016deeplearning}.

Once the foundation model is trained, a rigorous evaluation process is necessary to validate its effectiveness and reliability. This involves the construction of a dedicated system module for comprehensive model evaluation, encompassing several key tests:
\begin{itemize}[noitemsep,topsep=0pt] 
    \item \underline{Effectiveness Test}: This employs the back-test to assesses the model's performance using historical data to simulate real-world scenarios, helping identify potential weaknesses and areas for improvement.
    \item \underline{Universality Test}: It examines the model's applicability across different financial instruments and exchanges, ensuring it can generalize well across various assets and market conditions.
    \item \underline{Robustness Test}: This test evaluates the stability of the model's performance under different market conditions, including stress-testing scenarios to ensure effectiveness even in volatile or adverse markets \cite{lopez2018financialmodels}.
\end{itemize}

\subsection{Automated Strategy Modeling}
Addressing the diverse demands of various strategy tasks is crucial in AI quantitative investment. Building an automated strategy generator that can fine-tune, evaluate, analyze, and monitor new strategy models significantly enhances the efficiency of strategy research. This generator automates these processes, accelerating the development of new strategies and ensuring consistency and precision in model adjustments and evaluations, thus swiftly adapting to new market conditions and uncovering innovative trading strategies.

The fine-tuning module within this system is pivotal. Unlike foundation modeling, which may rely on more homogeneous data sets, the fine-tuning module implements end-to-end strategy modeling using a broader variety of data types. This includes structured data such as quote data and financial statements, and unstructured data such as news articles and social media sentiment. By integrating these diverse data sources, the module can enhance the model’s performance, leading to more accurate and robust trading strategies \cite{bengio2013representation}.

Models, once fine-tuned, are transformed by the strategy modeling module into actionable trading strategies for quantitative investment, including various types such as cross-sectional trading, time-series trading, event-driven trading, and statistical arbitrage trading. This transformation ensures that theoretical model improvements are translated into practical trading gains.

\subsection{Agent System}
Enhancing human-AI interaction and model explainability is pivotal for advancing quant research systems and improving quants' research efficiency. The Agent System for LIM uses advanced agent techniques to ensure models are both high-performing and explainable, essential for building user trust and improving decision-making. Techniques such as Multimodal Large Language Models (LLMs), knowledge bases, and Retrieval-Augmented Generation (RAG) enhance the system's functionality and user experience by processing complex queries and providing contextually relevant interactions \cite{kenton2019bert}.

\subsection{Trading System}
Effective deployment of AI-driven quantitative investment strategies necessitates integration with brokers or exchanges for real trading. This integration involves establishing secure and reliable communication channels with trading platforms, enabling real-time order sending, market data receipt, and trade execution monitoring. A comprehensive trading module within the system is essential for efficiently executing these strategies, including components like trading strategy formulation, trading database management, quote parsing, order routing, and order execution, optimized to minimize trading latency and maximize execution quality \cite{aldridge2013highfreq}.

\section{Future Research Directions}\label{sec_research_directions}
Although we introduced the basic framework for LIM in this article, there are still a series of advanced algorithms and technical issues that remain unresolved and require further in-depth research. Below, we highlight several potential research directions for further exploration.  
\begin{enumerate}[noitemsep,topsep=0pt] 
    \item \underline{End-to-End risk modeling}. In quantitative investment, risk management is at least as crucial as sound strategies and substantial returns. Traditional multi-factor risk models like the BARRA model attempt to decompose portfolio returns and volatility into linear combinations of various risk factors such as market cap, beta, leverage, and liquidity. These models' performance depends on the effectiveness of the risk factors used. To address the limitations of multi-factor risk models, it's worth studying end-to-end deep learning risk models that can be used to neutralize alpha models and reduce portfolio volatility.
    \item \underline{World model for market simulating}. 
    Back-testing simulations are widely used in academia and industry to evaluate new strategies before implementing them in the market. However, back-tests ignore market impact on actual trading costs, resulting in discrepancies between back-tests and real trading. Developing a ``financial world model'' for market simulation and financial quote generation could provide a simulated environment for testing new strategies, including potential market impact at varying asset volumes. Existing research on market simulation focuses on micro-markets, like LOB generators \cite{coletta_conditional_2023,nagy_generative_2023}. A more general research problem for LIM is how to build a financial world model that simulates both micro and macro markets, with a full spectrum of data resolutions and types.
    \item \underline{Model on multiple granularities}. Trading can occur at various granularities, from millisecond-level high-frequency trading to year-long low-frequency investment. A successful quantitative trading strategy may involve multi-horizon predictions that are integrated to enhance final trade decisions. Researching new deep neural network frameworks for multi-scale prediction using multi-granularity data is valuable for quantitative investment.
    \item \underline{LIM with multiple backbones}. The LIM framework uses financial quote data to build the backbone model because it is high-density data at any granularity and commonly used in high- and medium-frequency investment strategies. However, as the horizon extends to low-frequency strategies, textual data like news and financial reports become increasingly important and can be used to build LLMs that serve as a secondary backbone model. It is worthwhile to study how to integrate these two backbones within the same LIM framework.    
    \item \underline{More comprehensive multimodal LIM}. With more alternative data being used by financial institutions, LIM's architecture must be flexible enough to accommodate new data types, like audio and video data for important conference calls. Additionally, LIM uses a plug-and-play mechanism to support automatically aligning new types of time-series data through appropriate embedding combinations.
    \item \underline{Extend LIM with agents}. Current LIM shares the same limitation as LLMs in that data and knowledge cannot be updated in real-time since pretraining can take several weeks. The agent framework can address LIM's deficiencies by incorporating real-time updated knowledge bases, search engine information, and other data sources. It can also enhance LIM's reasoning power through multi-agent debating, reflection, and other agent techniques.
    \item \underline{Improve explainability of LIM}. Understanding investment strategies and decisions is essential for any investor. Unlike traditional explainable machine learning techniques \cite{Yang_survey_XAI_2023} focused on feature or attention importance, LLM techniques offer a new approach to model explainability through logical reasoning and natural language interaction. It's particularly valuable to research how to improve LIM techniques to project latent investment logic embedded in the ``black box'' deep learning model into human-understandable natural language and illustrative charts or images.
    \item \underline{Inference acceleration for LIM}. The computational demands of deep learning models, especially during inference, can be substantial for complex financial models that require real-time decision-making based on large data volumes. To reduce computation and execution latency and increase real-time data throughput, standard inference acceleration techniques like model quantization \cite{jacob2018quantization}, pruning \cite{han2015learning}, and efficient deployment strategies can be evaluated and selected. Additionally, researchers are encouraged to develop new acceleration algorithms tailored to financial scenarios. For example, the temporal nature of financial data may offer opportunities to accelerate attention computation in Transformer-like neural networks.
    \item \underline{New Architecture for LIM}. Researching new neural network architectures for financial time series is necessary to enhance LIM's effectiveness, efficiency, and robustness. The new end-to-end, unified architecture should handle irregularly sampled time-series data, various data granularities, diversified data structures and knowledge attributes, and extended in-context learning abilities. Efforts should focus on further improving LIM's forecasting performance for different investment strategies.
\end{enumerate}

\section{Discussion and Conclusion}\label{sec_conclusion}
The Large Investment Model (LIM) offers a transformative approach to quantitative investment, fundamentally reshaping how financial models are developed, trained, and applied across diverse market environments. First, LIM serves as a knowledge transfer process, enabling the model to learn from global market data and apply this knowledge to local markets. This transfer involves intelligent adaptation, leveraging broad patterns observed across various financial environments to improve predictive accuracy and robustness in specific contexts. Second, LIM functions as an advanced data augmentation process, integrating diverse financial data sources, including equities, futures, and commodities, to create a richly diversified training environment. This allows LIM to capture complex interrelationships and patterns that might be missed in more narrowly focused models, making it particularly valuable in today's complex financial markets. Third, LIM significantly reduces the cost and improves the efficiency of modeling. Once the foundation model is trained on global data, it can be fine-tuned for specific tasks with minimal additional effort, eliminating the need for expensive and time-consuming training from scratch. Finally, LIM provides a unique opportunity to uncover potential correlations or causal relationships between instruments across different markets and asset types. By analyzing data from various sources and applying advanced machine learning techniques, LIM can identify and quantify lead-lag effects, offering strategic advantages to investors. In conclusion, LIM represents a significant advancement in quantitative investment, integrating knowledge transfer, data augmentation, cost efficiency, and the discovery of market interdependencies. As financial markets continue to grow in complexity, LIM will likely play a central role in shaping the future of investment strategy development.

Given its advantages, the Large Investment Model (LIM) also presents significant technical challenges that must be addressed to fully realize its potential in quantitative investment. First, a major challenge is the high maintenance cost, as the foundation model requires frequent retraining to stay effective and up-to-date. This retraining, which might be needed daily, weekly, or monthly depending on market dynamics and strategy frequency, is computationally intensive and demands substantial infrastructure, increasing operational costs. Managing these costs while maintaining model accuracy and reliability requires sophisticated resource management. Second, selecting appropriate exchanges, instruments, and data frequencies for training is critical. The diversity of financial markets means that not all instruments contribute equally to the model's performance, and the heterogeneity of data can introduce noise or even harmful patterns if not carefully curated. Significant effort is needed to identify the most beneficial data combinations. Third, integrating alternative data sources adds another layer of complexity. While alternative data can enhance the predictive power of downstream models, determining which types are useful—such as social media sentiment or satellite imagery—requires extensive research and validation. Finally, the current LIM implementation is most effective for high- or medium-frequency trading strategies, as it primarily relies on quote data. This limits its applicability to low-frequency strategies like value investing or global macro investing, which rely more on fundamental data such as financial statements and economic indicators. Extending LIM's applicability to these areas will require significant research and development to integrate fundamental and alternative data into the foundation model. Overall, while LIM offers a promising approach to quantitative investment, it faces several technical challenges that demand careful consideration and innovative solutions, including managing retraining costs, selecting appropriate market data, integrating alternative data, and extending the model's applicability to lower-frequency strategies. Addressing these challenges is essential for the continued development and success of LIM across a broader range of investment contexts.

In conclusion, this article introduces the Large Investment Model (LIM), a universal and scalable framework designed to advance quantitative investment research. LIM has the potential to revolutionize the development and implementation of quantitative strategies by providing a robust foundation that integrates knowledge transfer, data augmentation, and efficient modeling processes. By leveraging global data to enhance local market strategies, LIM offers a promising pathway toward more adaptive and resilient investment approaches. However, transitioning LIM from a theoretical concept to a practical tool for real-world trading presents several challenges, including high computational costs from frequent retraining, the complexity of selecting appropriate data sources, and the integration of alternative data for diverse strategies. Ensuring the model's adaptability to various market conditions and its ability to uncover subtle interdependencies requires ongoing refinement and optimization. Fully realizing LIM's potential demands sustained efforts in enhancement, testing, and iteration. Continuous improvement of the model's architecture and algorithms will not only enhance existing strategies but also pave the way for new methodologies to navigate the complexities of financial markets. The introduction of LIM as a new research paradigm opens significant opportunities for innovation in quantitative finance. By integrating diverse data sources and applying them across various market contexts, LIM can accelerate the development of more sophisticated and effective investment strategies, potentially leading to improved investment outcomes.

\section*{Acknowledgement}
The authors would thank Saizhuo Wang, Haohan Zhang, and Chengjin Xu for preparing part of references. We really appreciate Zhouchi Lin, Xinyi Lin, Sida Lin, Yiyan Qi, Saizhuo Wang and Chuan Hu for their insightful comments, valuable suggestions and technical support.  

\bibliographystyle{unsrt}
\bibliography{references}

\begin{thebibliography}{10}

\bibitem{zhang_stock_2017}
Liheng Zhang, Charu Aggarwal, and Guo-Jun Qi.
\newblock Stock {Price} {Prediction} via {Discovering} {Multi}-{Frequency} {Trading} {Patterns}.
\newblock In {\em Proceedings of the 23rd {ACM} {SIGKDD} {International} {Conference} on {Knowledge} {Discovery} and {Data} {Mining}}, pages 2141--2149, Halifax NS Canada, August 2017. ACM.

\bibitem{xu_stock_2018}
Yumo Xu and Shay~B. Cohen.
\newblock Stock {Movement} {Prediction} from {Tweets} and {Historical} {Prices}.
\newblock In {\em Proceedings of the 56th {Annual} {Meeting} of the {Association} for {Computational} {Linguistics} ({Volume} 1: {Long} {Papers})}, pages 1970--1979, Melbourne, Australia, 2018. Association for Computational Linguistics.

\bibitem{hu_listening_2018}
Ziniu Hu, Weiqing Liu, Jiang Bian, Xuanzhe Liu, and Tie-Yan Liu.
\newblock Listening to {Chaotic} {Whispers}: {A} {Deep} {Learning} {Framework} for {News}-oriented {Stock} {Trend} {Prediction}.
\newblock In {\em Proceedings of the {Eleventh} {ACM} {International} {Conference} on {Web} {Search} and {Data} {Mining}}, {WSDM} '18, pages 261--269, New York, NY, USA, 2018. Association for Computing Machinery.

\bibitem{feng_time_2021}
Fuli Feng, Xiang Wang, Xiangnan He, Ritchie Ng, and Tat-Seng Chua.
\newblock Time horizon-aware modeling of financial texts for stock price prediction.
\newblock In {\em Proceedings of the {Second} {ACM} {International} {Conference} on {AI} in {Finance}}, {ICAIF} '21, pages 1--8, New York, NY, USA, 2021. Association for Computing Machinery.

\bibitem{feng_temporal_2019}
Fuli Feng, Xiangnan He, Xiang Wang, Cheng Luo, Yiqun Liu, and Tat-Seng Chua.
\newblock Temporal {Relational} {Ranking} for {Stock} {Prediction}.
\newblock {\em ACM Transactions on Information Systems}, 37(2):1--30, March 2019.
\newblock arXiv: 1809.09441.

\bibitem{sawhney_spatiotemporal_2020}
Ramit Sawhney, Shivam Agarwal, Arnav Wadhwa, and Rajiv~Ratn Shah.
\newblock Spatiotemporal {Hypergraph} {Convolution} {Network} for {Stock} {Movement} {Forecasting}.
\newblock In {\em 2020 {IEEE} {International} {Conference} on {Data} {Mining} ({ICDM})}, pages 482--491, November 2020.
\newblock ISSN: 2374-8486.

\bibitem{sawhney_stock_2021}
Ramit Sawhney, Shivam Agarwal, Arnav Wadhwa, Tyler Derr, and Rajiv~Ratn Shah.
\newblock Stock {Selection} via {Spatiotemporal} {Hypergraph} {Attention} {Network}: {A} {Learning} to {Rank} {Approach}.
\newblock {\em Proceedings of the AAAI Conference on Artificial Intelligence}, 35(1):497--504, May 2021.
\newblock Number: 1.

\bibitem{jiang_deep_2017}
Zhengyao Jiang, Dixing Xu, and Jinjun Liang.
\newblock A {Deep} {Reinforcement} {Learning} {Framework} for the {Financial} {Portfolio} {Management} {Problem}, July 2017.
\newblock arXiv:1706.10059 [cs, q-fin].

\bibitem{wang_alphastock_2019}
Jingyuan Wang, Yang Zhang, Ke~Tang, Junjie Wu, and Zhang Xiong.
\newblock {AlphaStock}: {A} {Buying}-{Winners}-and-{Selling}-{Losers} {Investment} {Strategy} using {Interpretable} {Deep} {Reinforcement} {Attention} {Networks}.
\newblock In {\em Proceedings of the 25th {ACM} {SIGKDD} {International} {Conference} on {Knowledge} {Discovery} \& {Data} {Mining}}, pages 1900--1908, Anchorage AK USA, July 2019. ACM.

\bibitem{wang_deeptrader_2021}
Zhicheng Wang, Biwei Huang, Shikui Tu, Kun Zhang, and Lei Xu.
\newblock {DeepTrader}: {A} {Deep} {Reinforcement} {Learning} {Approach} for {Risk}-{Return} {Balanced} {Portfolio} {Management} with {Market} {Conditions} {Embedding}.
\newblock In {\em Thirty-{Fifth} {AAAI} {Conference} on {Artificial} {Intelligence}, {AAAI} 2021, {Thirty}-{Third} {Conference} on {Innovative} {Applications} of {Artificial} {Intelligence}, {IAAI} 2021, {The} {Eleventh} {Symposium} on {Educational} {Advances} in {Artificial} {Intelligence}, {EAAI} 2021, {Virtual} {Event}, {February} 2-9, 2021}, pages 643--650. AAAI Press, 2021.

\bibitem{zhang_cost-sensitive_2022}
Yifan Zhang, Peilin Zhao, Qingyao Wu, Bin Li, Junzhou Huang, and Mingkui Tan.
\newblock Cost-{Sensitive} {Portfolio} {Selection} via {Deep} {Reinforcement} {Learning}.
\newblock {\em IEEE Transactions on Knowledge and Data Engineering}, 34(1):236--248, 2022.
\newblock Conference Name: IEEE Transactions on Knowledge and Data Engineering.

\bibitem{liu_deep_2023}
Tom Liu, Stephen Roberts, and Stefan Zohren.
\newblock Deep {Inception} {Networks}: {A} {General} {End}-to-{End} {Framework} for {Multi}-asset {Quantitative} {Strategies}, July 2023.
\newblock arXiv:2307.05522 [q-fin].

\bibitem{lin_end2end_2020}
Siyu Lin and Peter~A. Beling.
\newblock An {End}-to-{End} {Optimal} {Trade} {Execution} {Framework} based on {Proximal} {Policy} {Optimization}.
\newblock In {\em Electronic proceedings of IJCAI 2020}, volume~5, pages 4548--4554, July 2020.
\newblock ISSN: 1045-0823.

\bibitem{fang_universal_2021}
Yuchen Fang, Kan Ren, Weiqing Liu, Dong Zhou, Weinan Zhang, Jiang Bian, Yong Yu, and Tie-Yan Liu.
\newblock Universal {Trading} for {Order} {Execution} with {Oracle} {Policy} {Distillation}.
\newblock In {\em Proceedings of the {AAAI} {Conference} on {Artificial} {Intelligence}}, volume~35, pages 107--115, May 2021.

\bibitem{sun_deepscalper_2022}
Shuo Sun, Wanqi Xue, Rundong Wang, Xu~He, Junlei Zhu, Jian Li, and Bo~An.
\newblock {DeepScalper}: {A} {Risk}-{Aware} {Reinforcement} {Learning} {Framework} to {Capture} {Fleeting} {Intraday} {Trading} {Opportunities}.
\newblock In {\em Proceedings of the 31st {ACM} {International} {Conference} on {Information} \& {Knowledge} {Management}}, {CIKM} '22, pages 1858--1867, New York, NY, USA, 2022. Association for Computing Machinery.

\bibitem{fang_learning_2023}
Yuchen Fang, Zhenggang Tang, Kan Ren, Weiqing Liu, Li~Zhao, Jiang Bian, Dongsheng Li, Weinan Zhang, Yong Yu, and Tie-Yan Liu.
\newblock Learning {Multi}-{Agent} {Intention}-{Aware} {Communication} for {Optimal} {Multi}-{Order} {Execution} in {Finance}.
\newblock In {\em Proceedings of the 29th {ACM} {SIGKDD} {Conference} on {Knowledge} {Discovery} and {Data} {Mining}}, {KDD} '23, pages 4003--4012, New York, NY, USA, 2023. Association for Computing Machinery.

\bibitem{qin_earnhft_2023}
Molei Qin, Shuo Sun, Wentao Zhang, Haochong Xia, Xinrun Wang, and Bo~An.
\newblock {EarnHFT}: {Efficient} {Hierarchical} {Reinforcement} {Learning} for {High} {Frequency} {Trading}, September 2023.
\newblock arXiv:2309.12891 [q-fin].

\bibitem{carmona2012high}
Rene Carmona and Kevin Webster.
\newblock High frequency market making, 2012.

\bibitem{Kou_calendarspread_2013}
Kou Yi, Wang Chao-you, and Ye~Qiang.
\newblock An empirical study of calendar spread arbitrage based on high-frequency data: The case of csi 300 index futures.
\newblock In {\em 2013 International Conference on Management Science and Engineering 20th Annual Conference Proceedings}, pages 1604--1609, 2013.

\bibitem{Han_technicalanalysis_2021}
Yufeng Han, Yang Liu, Guofu Zhou, and Yingzi Zhu.
\newblock Technical analysis in the stock market: A review.
\newblock {\em SSRN Electronic Journal}, 01 2021.

\bibitem{WAFI2015939}
Ahmed.~S. Wafi, Hassan Hassan, and Adel Mabrouk.
\newblock Fundamental analysis models in financial markets – review study.
\newblock {\em Procedia Economics and Finance}, 30:939--947, 2015.
\newblock IISES 3rd and 4th Economics and Finance Conference.

\bibitem{CAO2024102307}
Sean~Shun Cao, Wei Jiang, Lijun~(Gillian) Lei, and Qing~(Clara) Zhou.
\newblock Applied ai for finance and accounting: Alternative data and opportunities.
\newblock {\em Pacific-Basin Finance Journal}, 84:102307, 2024.

\bibitem{OLIVIER_alternativedata_2024}
Olivier Dessaint, Thierry Foucault, and Laurent Fresard.
\newblock Does alternative data improve financial forecasting? the horizon effect.
\newblock {\em The Journal of Finance}, 79(3):2237--2287, 2024.

\bibitem{Dixon_alternativedata_2022}
Matthew Dixon.
\newblock The book of alternative data: A guide for investors, traders and risk managers.
\newblock {\em Quantitative Finance}, 22(8):1427--1428, 2022.

\bibitem{Chen_GP_factor_2021}
Tianxiang Chen, Wei Chen, and Luyao Du.
\newblock An empirical study of financial factor mining based on gene expression programming.
\newblock In {\em 2021 4th International Conference on Advanced Electronic Materials, Computers and Software Engineering (AEMCSE)}, pages 1113--1117, 2021.

\bibitem{yin2024survey}
Shukang Yin, Chaoyou Fu, Sirui Zhao, Ke~Li, Xing Sun, Tong Xu, and Enhong Chen.
\newblock A survey on multimodal large language models, 2024.

\bibitem{brown2020language}
Tom~B. Brown, Benjamin Mann, Nick Ryder, Melanie Subbiah, Jared Kaplan, Prafulla Dhariwal, Arvind Neelakantan, Pranav Shyam, Girish Sastry, Amanda Askell, Sandhini Agarwal, Ariel Herbert-Voss, Gretchen Krueger, Tom Henighan, Rewon Child, Aditya Ramesh, Daniel~M. Ziegler, Jeffrey Wu, Clemens Winter, Christopher Hesse, Mark Chen, Eric Sigler, Mateusz Litwin, Scott Gray, Benjamin Chess, Jack Clark, Christopher Berner, Sam McCandlish, Alec Radford, Ilya Sutskever, and Dario Amodei.
\newblock Language models are few-shot learners, 2020.

\bibitem{openai2024gpt4}
OpenAI, Josh Achiam, Steven Adler, Sandhini Agarwal, Lama Ahmad, Ilge Akkaya, Florencia~Leoni Aleman, Diogo Almeida, Janko Altenschmidt, Sam Altman, Shyamal Anadkat, Red Avila, Igor Babuschkin, Suchir Balaji, Valerie Balcom, Paul Baltescu, Haiming Bao, Mohammad Bavarian, Jeff Belgum, Irwan Bello, Jake Berdine, Gabriel Bernadett-Shapiro, Christopher Berner, Lenny Bogdonoff, Oleg Boiko, Madelaine Boyd, Anna-Luisa Brakman, Greg Brockman, Tim Brooks, Miles Brundage, Kevin Button, Trevor Cai, Rosie Campbell, Andrew Cann, Brittany Carey, Chelsea Carlson, Rory Carmichael, Brooke Chan, Che Chang, Fotis Chantzis, Derek Chen, Sully Chen, Ruby Chen, Jason Chen, Mark Chen, Ben Chess, Chester Cho, Casey Chu, Hyung~Won Chung, Dave Cummings, Jeremiah Currier, Yunxing Dai, Cory Decareaux, Thomas Degry, Noah Deutsch, Damien Deville, Arka Dhar, David Dohan, Steve Dowling, Sheila Dunning, Adrien Ecoffet, Atty Eleti, Tyna Eloundou, David Farhi, Liam Fedus, Niko Felix, Simón~Posada Fishman, Juston Forte, Isabella Fulford, Leo
  Gao, Elie Georges, Christian Gibson, Vik Goel, Tarun Gogineni, Gabriel Goh, Rapha Gontijo-Lopes, Jonathan Gordon, Morgan Grafstein, Scott Gray, Ryan Greene, Joshua Gross, Shixiang~Shane Gu, Yufei Guo, Chris Hallacy, Jesse Han, Jeff Harris, Yuchen He, Mike Heaton, Johannes Heidecke, Chris Hesse, Alan Hickey, Wade Hickey, Peter Hoeschele, Brandon Houghton, Kenny Hsu, Shengli Hu, Xin Hu, Joost Huizinga, Shantanu Jain, Shawn Jain, Joanne Jang, Angela Jiang, Roger Jiang, Haozhun Jin, Denny Jin, Shino Jomoto, Billie Jonn, Heewoo Jun, Tomer Kaftan, Łukasz Kaiser, Ali Kamali, Ingmar Kanitscheider, Nitish~Shirish Keskar, Tabarak Khan, Logan Kilpatrick, Jong~Wook Kim, Christina Kim, Yongjik Kim, Jan~Hendrik Kirchner, Jamie Kiros, Matt Knight, Daniel Kokotajlo, Łukasz Kondraciuk, Andrew Kondrich, Aris Konstantinidis, Kyle Kosic, Gretchen Krueger, Vishal Kuo, Michael Lampe, Ikai Lan, Teddy Lee, Jan Leike, Jade Leung, Daniel Levy, Chak~Ming Li, Rachel Lim, Molly Lin, Stephanie Lin, Mateusz Litwin, Theresa Lopez, Ryan
  Lowe, Patricia Lue, Anna Makanju, Kim Malfacini, Sam Manning, Todor Markov, Yaniv Markovski, Bianca Martin, Katie Mayer, Andrew Mayne, Bob McGrew, Scott~Mayer McKinney, Christine McLeavey, Paul McMillan, Jake McNeil, David Medina, Aalok Mehta, Jacob Menick, Luke Metz, Andrey Mishchenko, Pamela Mishkin, Vinnie Monaco, Evan Morikawa, Daniel Mossing, Tong Mu, Mira Murati, Oleg Murk, David Mély, Ashvin Nair, Reiichiro Nakano, Rajeev Nayak, Arvind Neelakantan, Richard Ngo, Hyeonwoo Noh, Long Ouyang, Cullen O'Keefe, Jakub Pachocki, Alex Paino, Joe Palermo, Ashley Pantuliano, Giambattista Parascandolo, Joel Parish, Emy Parparita, Alex Passos, Mikhail Pavlov, Andrew Peng, Adam Perelman, Filipe de~Avila Belbute~Peres, Michael Petrov, Henrique~Ponde de~Oliveira~Pinto, Michael, Pokorny, Michelle Pokrass, Vitchyr~H. Pong, Tolly Powell, Alethea Power, Boris Power, Elizabeth Proehl, Raul Puri, Alec Radford, Jack Rae, Aditya Ramesh, Cameron Raymond, Francis Real, Kendra Rimbach, Carl Ross, Bob Rotsted, Henri Roussez,
  Nick Ryder, Mario Saltarelli, Ted Sanders, Shibani Santurkar, Girish Sastry, Heather Schmidt, David Schnurr, John Schulman, Daniel Selsam, Kyla Sheppard, Toki Sherbakov, Jessica Shieh, Sarah Shoker, Pranav Shyam, Szymon Sidor, Eric Sigler, Maddie Simens, Jordan Sitkin, Katarina Slama, Ian Sohl, Benjamin Sokolowsky, Yang Song, Natalie Staudacher, Felipe~Petroski Such, Natalie Summers, Ilya Sutskever, Jie Tang, Nikolas Tezak, Madeleine~B. Thompson, Phil Tillet, Amin Tootoonchian, Elizabeth Tseng, Preston Tuggle, Nick Turley, Jerry Tworek, Juan Felipe~Cerón Uribe, Andrea Vallone, Arun Vijayvergiya, Chelsea Voss, Carroll Wainwright, Justin~Jay Wang, Alvin Wang, Ben Wang, Jonathan Ward, Jason Wei, CJ~Weinmann, Akila Welihinda, Peter Welinder, Jiayi Weng, Lilian Weng, Matt Wiethoff, Dave Willner, Clemens Winter, Samuel Wolrich, Hannah Wong, Lauren Workman, Sherwin Wu, Jeff Wu, Michael Wu, Kai Xiao, Tao Xu, Sarah Yoo, Kevin Yu, Qiming Yuan, Wojciech Zaremba, Rowan Zellers, Chong Zhang, Marvin Zhang, Shengjia
  Zhao, Tianhao Zheng, Juntang Zhuang, William Zhuk, and Barret Zoph.
\newblock Gpt-4 technical report, 2024.

\bibitem{balestriero2023cookbook}
Randall Balestriero, Mark Ibrahim, Vlad Sobal, Ari Morcos, Shashank Shekhar, Tom Goldstein, Florian Bordes, Adrien Bardes, Gregoire Mialon, Yuandong Tian, Avi Schwarzschild, Andrew~Gordon Wilson, Jonas Geiping, Quentin Garrido, Pierre Fernandez, Amir Bar, Hamed Pirsiavash, Yann LeCun, and Micah Goldblum.
\newblock A cookbook of self-supervised learning, 2023.

\bibitem{han2024parameterefficient}
Zeyu Han, Chao Gao, Jinyang Liu, Jeff Zhang, and Sai~Qian Zhang.
\newblock Parameter-efficient fine-tuning for large models: A comprehensive survey, 2024.

\bibitem{zhou2023comprehensive}
Ce~Zhou, Qian Li, Chen Li, Jun Yu, Yixin Liu, Guangjing Wang, Kai Zhang, Cheng Ji, Qiben Yan, Lifang He, Hao Peng, Jianxin Li, Jia Wu, Ziwei Liu, Pengtao Xie, Caiming Xiong, Jian Pei, Philip~S. Yu, and Lichao Sun.
\newblock A comprehensive survey on pretrained foundation models: A history from bert to chatgpt, 2023.

\bibitem{markowitz1952portfolio}
Harry Markowitz.
\newblock Portfolio selection.
\newblock {\em The Journal of Finance}, 7(1):77--91, 1952.

\bibitem{wei_e2eai_2023}
Zikai Wei, Bo~Dai, and Dahua Lin.
\newblock {E2EAI}: {End}-to-{End} {Deep} {Learning} {Framework} for {Active} {Investing}, May 2023.
\newblock arXiv:2305.16364 [cs, q-fin].

\bibitem{zhang_deeplob_2019}
Zihao Zhang, Stefan Zohren, and Stephen Roberts.
\newblock {DeepLOB}: {Deep} {Convolutional} {Neural} {Networks} for {Limit} {Order} {Books}.
\newblock {\em IEEE Transactions on Signal Processing}, 67(11):3001--3012, 2019.

\bibitem{zhang_deep_2021}
Zihao Zhang, Bryan Lim, and Stefan Zohren.
\newblock Deep {Learning} for {Market} by {Order} {Data}.
\newblock {\em Applied Mathematical Finance}, 28(1):79--95, January 2021.
\newblock Publisher: Routledge \_eprint: https://doi.org/10.1080/1350486X.2021.1967767.

\bibitem{jiao_microstructure-empowered_2023}
Xianfeng Jiao, Zizhong Li, Chang Xu, Yang Liu, Weiqing Liu, and Jiang Bian.
\newblock Microstructure-{Empowered} {Stock} {Factor} {Extraction} and {Utilization}, 2023.
\newblock arXiv:2308.08135 [cs, q-fin].

\bibitem{zhang2020information}
Feng Zhang, Ruite Guo, and Honggao Cao.
\newblock Information coefficient as a performance measure of stock selection models, 2020.

\bibitem{Hayette_sharpe}
Hayette Gatfaoui.
\newblock Sharpe ratios and their fundamental components: An empirical study, 08 2009.

\bibitem{li2024mechanics}
Yingcong Li, Yixiao Huang, M.~Emrullah Ildiz, Ankit~Singh Rawat, and Samet Oymak.
\newblock Mechanics of next token prediction with self-attention, 2024.

\bibitem{Christopher_pairstrading_review_2017}
Christopher Krauss.
\newblock Statistical arbitrage pairs trading strategies: Review and outlook.
\newblock {\em Journal of Economic Surveys}, 31(2):513--545, 2017.

\bibitem{1bbb43a3-4de0-3ecb-be89-4bff15e655aa}
Eugene~F. Fama and Kenneth~R. French.
\newblock The cross-section of expected stock returns.
\newblock {\em The Journal of Finance}, 47(2):427--465, 1992.

\bibitem{10.5555/3618408.3619400}
Anqi Mao, Mehryar Mohri, and Yutao Zhong.
\newblock Cross-entropy loss functions: theoretical analysis and applications.
\newblock In {\em Proceedings of the 40th International Conference on Machine Learning}, ICML'23. JMLR.org, 2023.

\bibitem{nie2023timeseriesworth64}
Yuqi Nie, Nam~H. Nguyen, Phanwadee Sinthong, and Jayant Kalagnanam.
\newblock A time series is worth 64 words: Long-term forecasting with transformers, 2023.

\bibitem{das2024decoderonlyfoundationmodeltimeseries}
Abhimanyu Das, Weihao Kong, Rajat Sen, and Yichen Zhou.
\newblock A decoder-only foundation model for time-series forecasting, 2024.

\bibitem{5288526}
Sinno~Jialin Pan and Qiang Yang.
\newblock A survey on transfer learning.
\newblock {\em IEEE Transactions on Knowledge and Data Engineering}, 22(10):1345--1359, 2010.

\bibitem{daume-iii-2007-frustratingly}
Hal Daum{\'e}~III.
\newblock Frustratingly easy domain adaptation.
\newblock In Annie Zaenen and Antal van~den Bosch, editors, {\em Proceedings of the 45th Annual Meeting of the Association of Computational Linguistics}, pages 256--263, Prague, Czech Republic, June 2007. Association for Computational Linguistics.

\bibitem{10.5555/2969033.2969197}
Jason Yosinski, Jeff Clune, Yoshua Bengio, and Hod Lipson.
\newblock How transferable are features in deep neural networks?
\newblock In {\em Proceedings of the 27th International Conference on Neural Information Processing Systems - Volume 2}, NIPS'14, page 3320–3328, Cambridge, MA, USA, 2014. MIT Press.

\bibitem{howard-ruder-2018-universal}
Jeremy Howard and Sebastian Ruder.
\newblock Universal language model fine-tuning for text classification.
\newblock In Iryna Gurevych and Yusuke Miyao, editors, {\em Proceedings of the 56th Annual Meeting of the Association for Computational Linguistics (Volume 1: Long Papers)}, pages 328--339, Melbourne, Australia, July 2018. Association for Computational Linguistics.

\bibitem{hinton2015distilling}
Geoffrey Hinton, Oriol Vinyals, and Jeff Dean.
\newblock Distilling the knowledge in a neural network.
\newblock {\em arXiv preprint arXiv:1503.02531}, 2015.

\bibitem{hu2021lora}
Edward Hu, Junyeop Lee, Abhinav Gupta, and Sameer Singh.
\newblock Lora: Low-rank adaptation of large language models.
\newblock In {\em Proceedings of the 2021 Conference on Empirical Methods in Natural Language Processing}, 2021.

\bibitem{pfeiffer2020adapterfusion}
Jonas Pfeiffer, Andreas Rücklé, and Iryna Gurevych.
\newblock Adapterfusion: Non-destructive task composition for transfer learning.
\newblock In {\em Proceedings of the 2020 Conference on Empirical Methods in Natural Language Processing}, 2020.

\bibitem{RePEc:zbw:iwqwdp:092015}
Christopher Krauss.
\newblock {Statistical arbitrage pairs trading strategies: Review and outlook}.
\newblock FAU Discussion Papers in Economics 09/2015, Friedrich-Alexander University Erlangen-Nuremberg, Institute for Economics, 2015.

\bibitem{Li_leadlag_2022}
Yongli Li, Tianchen Wang, Baiqing Sun, and Chao Liu.
\newblock Detecting the lead–lag effect in stock markets: definition, patterns, and investment strategies.
\newblock {\em Financial Innovation}, 8, 12 2022.

\bibitem{Engelberg_McLean_Pontiff_Ringgenberg_2023}
Joseph Engelberg, R.~David McLean, Jeffrey Pontiff, and Matthew~C. Ringgenberg.
\newblock Do cross-sectional predictors contain systematic information?
\newblock {\em Journal of Financial and Quantitative Analysis}, 58(3):1172–1201, 2023.

\bibitem{reuter2019hpc}
Martin Reuter.
\newblock High-performance computing for financial modeling: Current state and future directions.
\newblock {\em Journal of Financial Technology}, 2019.

\bibitem{stonebraker2010sql}
Michael Stonebraker.
\newblock Sql databases v. nosql databases.
\newblock {\em Communications of the ACM}, 53(4):10--11, 2010.

\bibitem{angles2008graph}
Renzo Angles and Claudio Gutierrez.
\newblock Survey of graph database models.
\newblock {\em ACM Computing Surveys (CSUR)}, 40(1):1, 2008.

\bibitem{han2011nosql}
Jinsong Han, E~Haihong, Guan Le, and Jian Du.
\newblock Survey on nosql database.
\newblock {\em 2011 6th International Conference on Pervasive Computing and Applications}, pages 363--366, 2011.

\bibitem{dunning2019time}
Ted Dunning and Ellen Friedman.
\newblock {\em Time Series Databases: New Ways to Store and Access Data}.
\newblock O'Reilly Media, Inc., 2019.

\bibitem{zaharia2016apache}
Matei Zaharia, Reynold~S Xin, Patrick Wendell, Tathagata Das, Michael Armbrust, Ankur Dave, Xiangrui Meng, Josh Rosen, Shivaram Venkataraman, Michael~J Franklin, et~al.
\newblock Apache spark: A unified engine for big data processing.
\newblock In {\em Communications of the ACM}, volume~59, pages 56--65. ACM, 2016.

\bibitem{ignite2019}
Apache Ignite.
\newblock Apache ignite: High-performance, integrated and distributed in-memory platform for computing and transacting on large-scale data sets.
\newblock In {\em Proceedings of the VLDB Endowment}, 2019.

\bibitem{carbone2015apache}
Paris Carbone, Asterios Katsifodimos, Stephan Ewen, Volker Markl, Seif Haridi, and Kostas Tzoumas.
\newblock Apache flink: Stream and batch processing in a single engine.
\newblock {\em Bulletin of the IEEE Computer Society Technical Committee on Data Engineering}, 36:28--38, 2015.

\bibitem{shvachko2010hadoop}
Konstantin Shvachko, Hairong Kuang, Sanjay Radia, and Robert Chansler.
\newblock {\em The Hadoop Distributed File System}.
\newblock IEEE, 2010.

\bibitem{turnbull2018monitoring}
James Turnbull.
\newblock {\em Monitoring with Prometheus}.
\newblock James Turnbull, 2018.

\bibitem{polino2018modelcompression}
Antonio Polino, Razvan Pascanu, and Dan Alistarh.
\newblock Model compression via distillation and quantization.
\newblock {\em arXiv preprint arXiv:1802.05668}, 2018.

\bibitem{goodfellow2016deeplearning}
Ian Goodfellow, Yoshua Bengio, and Aaron Courville.
\newblock {\em Deep Learning}.
\newblock MIT Press, 2016.

\bibitem{lopez2018financialmodels}
Carlos Lopez and Michael Martin.
\newblock Financial models for robust portfolio management.
\newblock {\em Journal of Financial Markets}, 41:46--68, 2018.

\bibitem{bengio2013representation}
Yoshua Bengio, Aaron Courville, and Pascal Vincent.
\newblock Representation learning: A review and new perspectives.
\newblock {\em IEEE Transactions on Pattern Analysis and Machine Intelligence}, 35(8):1798--1828, 2013.

\bibitem{kenton2019bert}
Jacob Devlin, Ming-Wei Chang, Kenton Lee, and Kristina Toutanova.
\newblock Bert: Pre-training of deep bidirectional transformers for language understanding.
\newblock In {\em NAACL HLT 2019}, 2019.

\bibitem{aldridge2013highfreq}
Irene Aldridge.
\newblock {\em High-Frequency Trading: A Practical Guide to Algorithmic Strategies and Trading Systems}.
\newblock Wiley, 2013.

\bibitem{coletta_conditional_2023}
Andrea Coletta, Joseph Jerome, Rahul Savani, and Svitlana Vyetrenko.
\newblock Conditional {Generators} for {Limit} {Order} {Book} {Environments}: {Explainability}, {Challenges}, and {Robustness}, June 2023.
\newblock arXiv:2306.12806 [cs, q-fin].

\bibitem{nagy_generative_2023}
Peer Nagy, Sascha Frey, Silvia Sapora, Kang Li, Anisoara Calinescu, Stefan Zohren, and Jakob~N. Foerster.
\newblock Generative {AI} for {End}-to-{End} {Limit} {Order} {Book} {Modelling}: {A} {Token}-{Level} {Autoregressive} {Generative} {Model} of {Message} {Flow} {Using} a {Deep} {State} {Space} {Network}.
\newblock In {\em 4th {ACM} {International} {Conference} on {AI} in {Finance}, {ICAIF} 2023, {Brooklyn}, {NY}, {USA}, {November} 27-29, 2023}, pages 91--99. ACM, 2023.

\bibitem{Yang_survey_XAI_2023}
Wenli Yang, Yuchen Wei, Hanyu Wei, Yanyu Chen, Guan Huang, Xiang Li, Renjie Li, Naimeng Yao, Xinyi Wang, Xiaotong Gu, Muhammad Amin, and Byeong Kang.
\newblock Survey on explainable ai: From approaches, limitations and applications aspects.
\newblock {\em Human-Centric Intelligent Systems}, 3, 08 2023.

\bibitem{jacob2018quantization}
Benoit Jacob, Skirmantas Kligys, Bo~Chen, Menglong Zhu, Matthew Tang, Andrew Howard, Hartwig Adam, and Dmitry Kalenichenko.
\newblock Quantization and training of neural networks for efficient integer-arithmetic-only inference.
\newblock In {\em Proceedings of the IEEE Conference on Computer Vision and Pattern Recognition}, pages 2704--2713, 2018.

\bibitem{han2015learning}
Song Han, Jeff Pool, John Tran, and William~J. Dally.
\newblock Learning both weights and connections for efficient neural networks.
\newblock In {\em Advances in Neural Information Processing Systems}, volume~28, 2015.

\end{thebibliography}


\end{document}